\documentclass{emulateapj}

\setkeys{Gin}{width=\linewidth,height=0.9\textheight,keepaspectratio=true}

\newcommand{\Ha}{H$\alpha$}
\newcommand{\Hb}{H$\beta$}
\newcommand{\htwo}{H$_{2}$}
\newcommand{\Stwo}{S$_{2.12\mu}$}

\newcommand{\sbunits}{ergs s$^{-1}$ cm$^{-2}$ sr$^{-1}$}
\newcommand{\etal}{et~al.}
\newcommand{\kms}{km~s{$^{-1}$}}
\newcommand{\psec}{s{$^{-1}$}}

\newcommand{\cmq}{cm$^{-3}$}
\newcommand{\cms}{cm$^{-2}$}
\newcommand{\pers}{s$^{-1}$}

\newcommand{\vsun}{V$_{\hbox{$\odot$}}$}

\shorttitle{Knots in the Helix Nebula}
\shortauthors{O'Dell, Henney, and Ferland}

\received{2005 February 21}
\begin{document}


\title{A Multi-Instrument Study of the Helix Nebula Knots with the Hubble 
Space Telescope\altaffilmark{1,2}}

\altaffiltext{1}{Based in part on observations with the NASA/ESA Hubble
  Space Telescope, obtained at the Space Telescope Science Institute,
  which is operated by the Association of Universities for Research in
  Astronomy, Inc., under NASA Contract No. NAS 5-26555.}
\altaffiltext{2}{Based in part on observations obtained at the Cerro
  Tololo Interamerican Observa tory, which is operated by the
  Association of Universities for Research in Astronomy, Inc., under a
  Cooperative Agreement with the National Science Foundation.}


\author{C. R. O'Dell}
\affil{Department of Physics and Astronomy, Vanderbilt University,
Box 1807-B, Nashville, TN 37204}
\email{cr.odell@vanderbilt.edu}
\author{W. J. Henney\altaffilmark{3}}
\affil{Centro de Radioastronom\'{\i}a y Astrof\'{\i}sica, UNAM Campus Morelia,
  Apartado Postal 3-72, 58090 Morelia, Michoac\'an, M\'exico.}
\altaffiltext{3}{Work carried out in part while on sabbatical at
Department of Physics and Astronomy, University of Leeds, LS2~9JT,
UK.}
\and
\author{Gary J. Ferland}
\affil{Department of Physics, University of Kentucky, Lexington, KY 40506}




\begin{abstract}

We have conducted a combined observational and theoretical investigation of the ubiquitous knots in the 
Helix Nebula (NGC~7293). We have constructed a combined hydrodynamic+radiation model for the ionized portion of these knots and have accurately calculated a static model for their molecular regions.
Imaging observations in optical emission lines were made with the Hubble Space Telescope's
STIS spectrograph, operating in a ``slitless'' mode, complemented by WFPC2 images in several of the same
lines. The NICMOS camera was used to image the knots in \htwo. These observations, when combined with 
other studies of \htwo\ and CO provide a complete characterization of the knots. They possess dense
molecular cores of densities about 10$^{6}$ \cmq\ surrounded on the central star side by a zone of 
hot \htwo. The temperature of the \htwo\ emitting layer defies explanation either through detailed 
calculations for radiative equilibrium or for simplistic calculations for shock excitation. Further 
away from the core is the ionized zone, whose peculiar distribution of 
emission lines is explained by the expansion effects of material flowing through this region. The shadowed
region behind the core is the source of most of the CO emission from the knot and is of the low temperature expected for a radiatively heated molecular region.
\end{abstract}


\section{INTRODUCTION}

A pressing issue within the study of the planetary nebulae (PN) is the nature
of the highly compact, neutral core knots that have been found in all the nearby
PN studied by the Hubble Space Telescope (HST) and arguably are ubiquitous 
(O'Dell \etal\ 2002, henceforth O2002). Their origin is not established, although two extreme
models exist, the first, where they originate as knots in the extended atmosphere
of the precursor central star (Dyson \etal\ 1989, Hartquist \&\ Dyson 1997) and the second possibility is that they are
the results of instabilities occurring at the boundary between the ionized and
neutral zones within the nebulae (Capriotti 1973). Perhaps the most significant issue about them
is not their origin, rather ``What is their fate?'' This is because their survival
beyond the PN stage would mean that a large fraction of the material being 
put into the interstellar medium (ISM) by the PN phenomenon would be trapped in optically
thick knots, which would then become a new component of the ISM. However, to delve
into their origin or prognosticate their future, we must understand the objects as
they are now. Fortunately, these knots have become the subjects of recent observational
studies that probe from the outside to the inside of the knots and we can begin to
hope of having a complete picture of their nature. This paper presents new 
observational results for the nearest bright PN, NGC~7293--the Helix Nebula, that probe both the outer ionized layers and a portion of the
neutral inner core of its knots. These observations are then compared with new sophisticated 
models that accurately model the regions of origin of the emission and we present
a general model for the objects.

The Helix Nebula is a member of the polypolar class of PN where there is a smaller filled inner-disk
within a fainter outer-disk that is almost perpendicular to it (O'Dell, \etal\ 2004, henceforth
OMM2004). The outer-disk is open in the middle where it encloses the inner-disk and its opening is surrounded by a brighter feature called the outer-ring. Both are optically thick
to ionizing Lyman continuum (LyC) radiation. The innermost portions of the inner-disk contains a core of
He$^{++}$ emission, whose lack of
ions with high emissivity give the nebula a superficial appearance of having a
central cavity (O'Dell 1998, henceforth O1998).  The several thousand knots that are present in the Helix Nebula
begin to be seen about about the region of transition from He$^{++}$ to He$^{+}$ and
they are found with increasing concentration as one moves toward the ionization
boundary, both in the inner-disk and the outer-ring (OMM2004). 

The knots were originally discovered by Walter Baade and first reported
upon by Zanstra (1955) and Vorontzov-Velyaminov (1968). The next big step in the illumination of the
knots was the paper by Meaburn \etal\ 1992 (henceforth M1992), which established that the knots had highly 
ionized cusps of about 2\arcsec\ size facing the hot central star, little emission in the [O~III] emission
line, and that the dust in their central cores blocked out some of the background
nebular emission, the amount of this extinction providing information about their 
location within the nebula. This picture was extended by much higher spatial
resolution images made with the HST, which has now covered several portions of the
nebula in the northern quadrant (O'Dell \&\ Handron 1996, henceforth OH1996, O'Dell \& Burkert 1997, henceforth OB1997) and with lower signal to noise ratio, a
section in the main ring of emission to the northwest from the central star (O2002).
There are about 3,500 knots in the entire nebulae (OH1996) and the characteristic mass
of each is about 3 x 10$^{-5}$ M$\rm _{sun}$ (OB1997) or $\ga$ 1 x 10$^{-5}$ M$\rm _{sun}$ (Huggins \etal\ 2002, henceforth H2002), the former value meaning that the knots contain a total mass of about
0.1 M$\rm _{sun}$, which is about the same as all of the ionized gas. This indicates that the knots
represent an important ingredient in the mass-loss process of this PN.
The HST Wide Field and Planetary Camera 2 (WFPC2) allowed distinguishing between the
low ionization [N~II] emission and that of \Ha\ and [O~III].  This information, 
together with slitless spectra (O'Dell, \etal\ 2000, henceforth OHB2000) established that the ionization structure of the
bright knots was dissimilar to other photoionized structures in that the [N~II]
structure is more diffuse than that in \Ha, which could be explained by the rather
ad~hoc introduction of a peculiar electron temperature distribution, which is now
explained by the dynamical model introduced in this paper. The inner knots show 
well defined ``tails'' which has led to the knots sometimes being called ``cometary knots''. The radial alignment of these features has suggested to some authors that there may be a
radial outflow of material along them (Dyson, Hartquist, \&\ Biro 1993),
although the shadowing of ionizing Lyman Continuum radiation must play an
important role (Cant\'o \etal\ 1998, O'Dell 2000).

The regions within the bright ionized cusps have been probed by observations of
their molecular cores which now have sufficient resolution to 
make a clear delineation of the emission as arising from the core of the knots,
rather than an extended Photon Dominated Region (PDR) lying outside the ionization
front of the nebula. The entire Helix Nebula has been imaged (Speck \etal\ 2002,
henceforth S2002) in the \htwo\ 2.12 \micron\ line at a resolution of about 
2\arcsec. This study is supplemented by a 1.2\arcsec\ resolution image of one
of the knots by H2002.  Radio observations in other
molecules have been of progressively better spatial resolution, with the entire
nebula having now been mapped at 31\arcsec\ in CO by Young \etal\ 1999 (henceforth
Y1999) where the presence of multiple velocity components within the beam indicate that
there were emitting regions smaller than the beam size.
H2002 made CO observations with an elliptical beam of 7.9\arcsec x 3.8\arcsec\ 
of the same knot as in their high resolution \htwo\ study.
A splendid spectroscopic study at 6\arcsec\ spatial resolution of multiple
\htwo\ emission lines established (Cox \etal\ 1998) that the \htwo\ levels are in statistical equilibrium and that the temperature of the \htwo\ portions of
the cores is a surprising 900K. The best resolution study of neutral hydrogen 
is that of Rodr\'{\i}guez, Goss, \&\ Williams (2002) , although their 54.3\arcsec x 39.3\arcsec\ beam was 
insufficient to distinguish between emission coming from within the knots or the
PDR associated with the nebula's ionization front.

Fortunately, we know a lot of the basic characteristics of the system. The trigonometric
parallax of Harris \etal\ (1997) indicates that the distance is 213 parsecs, which means
that 1\arcsec\ = 3.19 x 10$^{15}$ cm and the 500\arcsec\ semimajor axis of the inner-disk (OMM2004)
corresponds to 0.52 pc.  The dynamic age of the inner-disk is 6,600 years (OMM2004). The 
central star has been measured accurately in the Far Ultra Violet (FUV, the ultraviolet flux with frequences below the Lyman limit) by Bohlin \etal\ (1982), who found an effective
temperature of 123,000 K. At 213 pc distance the star must have a bolometric 
luminosity of 120 L$\rm _{sun}$.   The central region has also been measured in the 
x-ray region from 0.1 to 2 KeV  (Leahy \etal\ 1994), who not only measured radiation
from the central star, but also found an additional
slightly extended source with a temperature of 8.7 x 10$^{6}$ K and a total
flux of 9 x 10$^{-14}$ ergs \cms\ \pers\, which converts to a total luminosity of the
non-stellar source of 1.3 x 10$^{-4}$ L$\rm _{sun}$.
There is no spectroscopic evidence (Cerruti-Sola \&\ Perinotto 1985) for continued outflow from the central
star, which is consistent with the late evolutionary stage of the star and the 
observation that there is not a central cavity in the nebular disk.

A note on nomenclature is in order. Various names have been used for the same features
in the multitude of papers addressing the Helix Nebula, this nomenclature often 
reflecting the background of the author and the state of knowledge. The compact 
features have been called filaments, globules, and cometary knots. In this paper
they will be called simply ``knots''. There have also been a variety of names applied
to the large scale bright structures within the bright ring seen in low ionization
line images and which lead to the original designation as the helix Nebula. In this
paper we will refer to these as ``loops''.  Whenever a fine-scale feature, such
as a knot, is to be designated specifically, we'll use the coordinate based system
introduced in OB1998, which avoids the confusion of serial or discovery-time based
naming systems.

We describe the new HST observations in \S~2. 
The observations are discussed in \S~3. In \S~4 we discuss these observations, showing that
the optical line properties require inclusion of the effects of advection and that no model
of the knots is able to explain the \htwo\ observations of this or other well-studied planetary
nebulae. 

\section{OBSERVATIONS}

The prime target of the observing program (GO-9489 in the Space Telescope Science
Institute, STScI, designation) was the isolated knot 378-801 (OB1997 system, C1 in 
Huggins \etal\ 1992, feature 1 in Figure 3a of Meaburn et al. 1998, henceforth M1998),
which lies nearly north of the central star and on the inside
edge of the ring of low ionization emission lines. It was  natural to select this
target because of the smooth background of the nebular emission, which makes it
easy to correct for that contaminating emission, and because it has been the subject
of multiple previous studies, including HST imaging (OH1996), HST low resolution 
spectroscopy (OHB2000), groundbased high resolution spectroscopy (M1998, O2002), and
the highest resolution imaging in CO and \htwo\ (H2002).  The knot lies in a direction
of position angle (PA) PA=356\arcdeg\ from the central star.

\subsection{Spectroscopic Observations}

\begin{figure}
\includegraphics{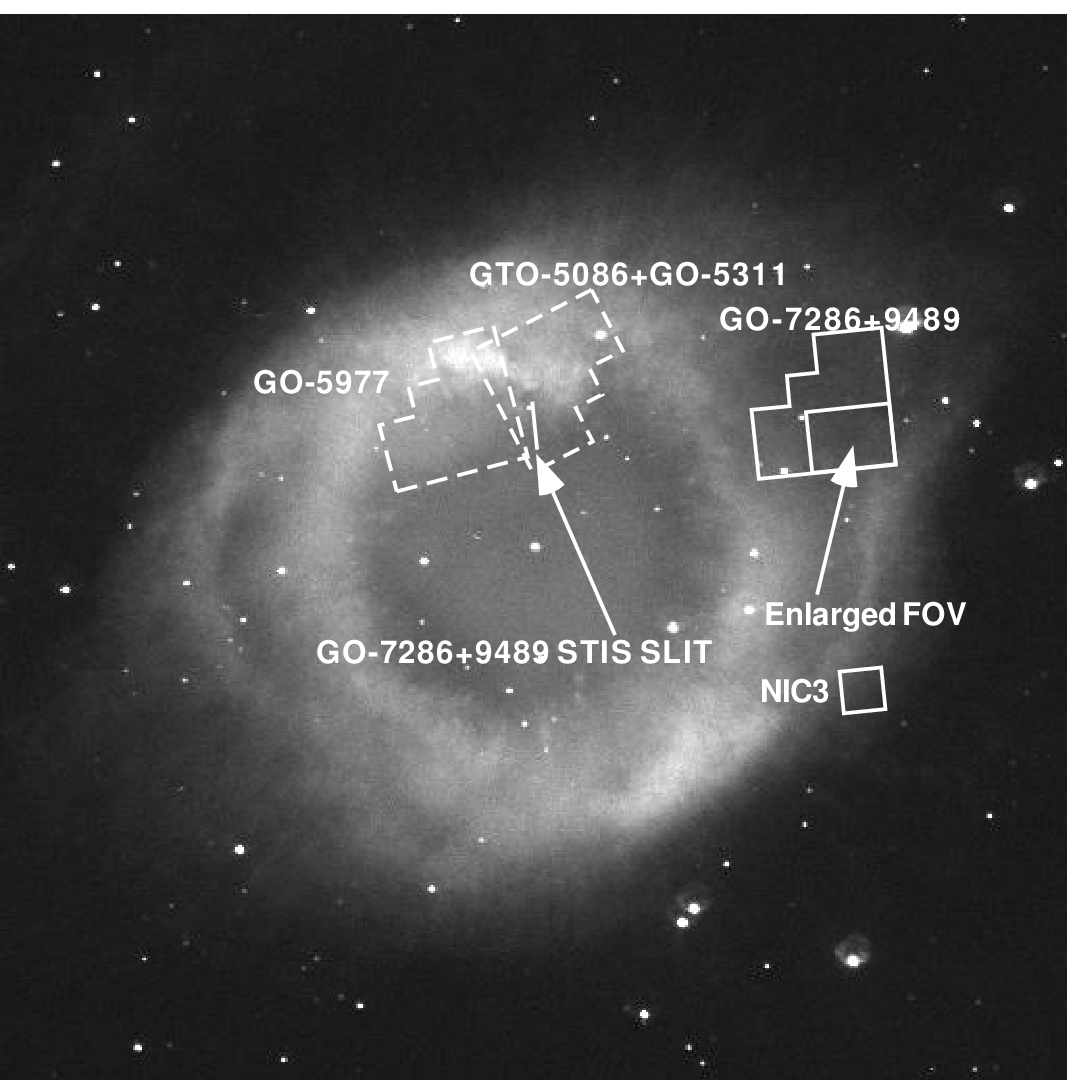}
\figcaption[2.1.A]{
This image of the Helix Nebula in \Hb\ (O1998) shows the location of the apertures
of the several data sets used in this study and HST program number that lead to
their creation.  This figure is similar to Figure 7 in O2002, except that the correct
position angle of the STIS slit (PA=6\arcdeg) is shown and the WFPC2 and NIC3 
coordinated parallel fields are shifted accordingly. The image is 1180\arcsec x
1180\arcsec\ and north is up.}

\end{figure}

\begin{figure}
\includegraphics{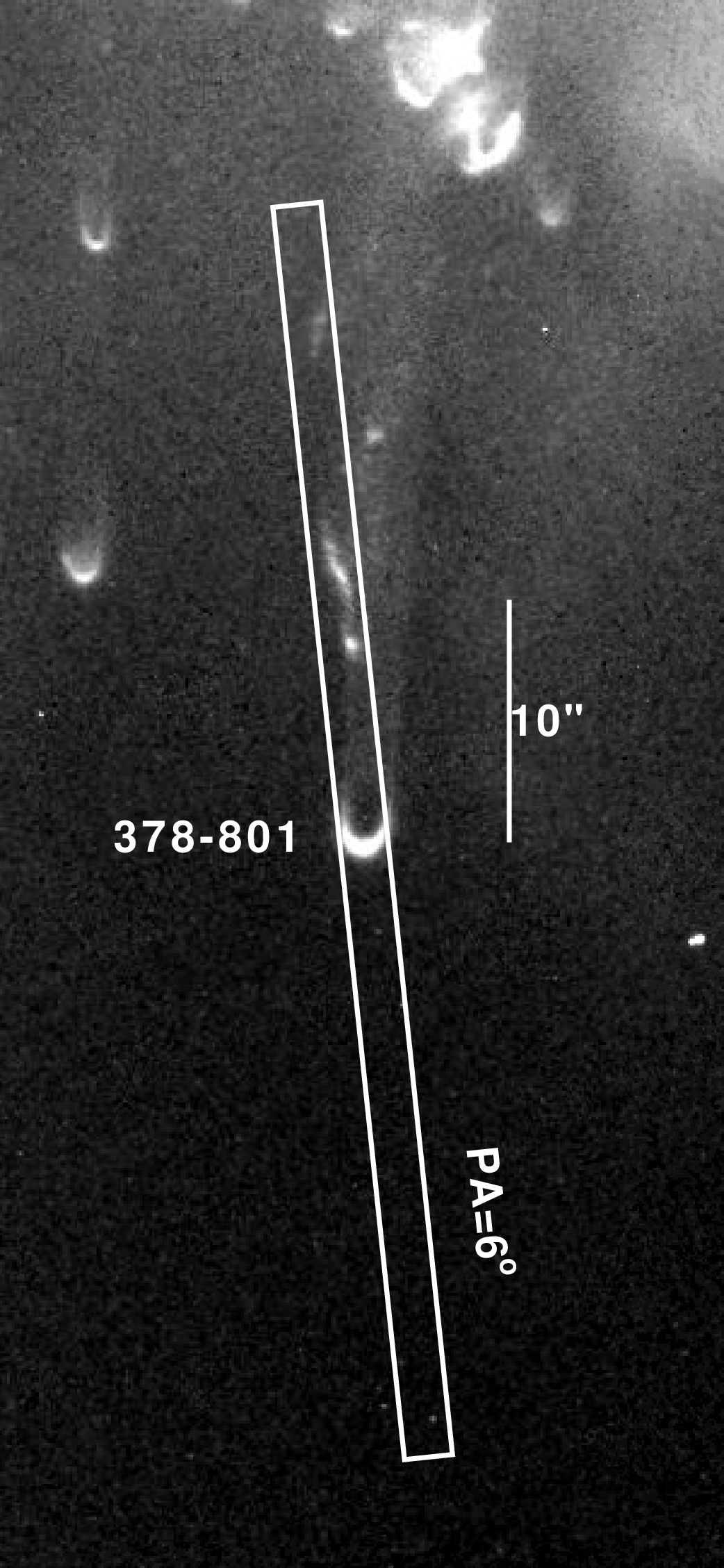}
\figcaption[2.1.B]{
This is a 30\arcsec x65\arcsec\ section of the sum of the GTO-5086 and GO-5311 
flux calibrated observations of \Ha\ and [N~II], with the knot 378-801 near the 
center.  The rectangle indicates the boundary and of the 2\arcsec x52\arcsec\ slit
used for the STIS observations discussed in this paper. North is up. The features
about 10\arcsec\ north of the bright cusp of 378-801 may be structure in the 
tail of this knot or may be a second knot that falls within the radiation shadow
of 378-801.}

\end{figure}

This knot was observed with the
Space Telescope Imaging Spectrometer (STIS,  Woodgate \etal\ 1998) through its 2\arcsec x 52\arcsec
entrance slit. Since the bright cusp of this knot is smaller than the width of the
entrance slit and the radiation is dominated by emission lines, STIS was able to 
form monochromatic images in the various lines, essentially functioning as a slitless
spectrograph. This pointing was the same as that employed in program GO-7286. The 
location of the entrance slit as projected on the nebula is shown in Figure 1
and Figure 2. Because of the paucity of candidate 
guidestars, the slit was oriented with PA=6\arcdeg, as were the observations in
program GO-7286.

Observations were made with multiple tilts of gratings G430M and G750M. A central
wavelength setting of 4961 \AA\ (2340 seconds exposure) gave useful observations of
the \Hb 4861 \AA, and [O~III] doublet 4959 \AA\ and 5007 \AA\ lines. A central wavelength
setting of 6252 \AA\ (4620 seconds exposure) gave useful observations of the [O~I]
doublet 6300 \AA\ and 6363 \AA. A central wavelength setting of 6581 \AA\ (4620 seconds)
gave useful observations of the same [O~I] lines, the \Ha\ line at 6563 \AA\ and the adjacent [N~II]
doublet 6548 \AA\ and 6583 \AA. These images were pipe-line processed by the STScI
and flux calibrated using their system that is tied to observations of standard 
stars through the same instrument configuration. The resulting images have pixel
sizes of 0.05\arcsec.
\subsection{Imaging Observations}
The wide field of good focus and large data storage capability of the HST allows
the making of parallel observations. WFPC2 (Holtzman \etal\ 1995) and NICMOS (Thompson \etal\ 1998) observations were made
during all of the primary target STIS observations. Since the primary target determines the
pointing of the spacecraft, one gets what is available in terms of the parallel
observations. However, the STIS slit could be placed in two orientations of 
180\arcdeg\ difference. We chose an orientation of the slit so that both the WFPC2
and NICMOS fields would fall onto the main body of the nebula.

The WFPC2 parallel field falls in a region outside of the main bright ring of
emission of the Helix and in the zone that may be the projection of the more distant
rotational axis of the nebula, which is a thick disk accompanied by perpendicular
low density plumes (O1998, M1998).

\subsubsection{WFPC2 Observations}
We made WFPC2 images of the field shown in Figure 1 in four filters. There were six
exposures in F656N for a total exposure time of 5600 seconds, six exposures in F658N
(6600 seconds total), four exposures in F502N (4000 seconds total), and three exposures in F547M (300 
seconds total). The multiple exposures were made to allow us to eliminate random events caused
by cosmic rays. The F502N filter primarily isolates [O~III] 5007 \AA\ emission, 
while F656N is dominated by \Ha\ 6563 \AA\ emission, and F658N is dominated by 
[N~II] 6583 \AA\ emission. However, each filter is affected by the underlying
continuum and the F656N and F658N filters are both affected by \Ha\ and [N~II]
emission, the latter becoming important in the innermost knots which have much
stronger [N~II] emission than \Ha\ emission. These observations were pipe-line
processed at the STScI, then combined with a similar set of observations made 
during program GO-7286, where the pointing was almost identical and the total exposure 
times were: F502N (4200 seconds), F547M (300 seconds), F656N (4600 seconds),
F658N (2200 seconds). These images were then rendered into monochromatic, calibrated
emission line images using the technique of O'Dell \&\ Doi (1998), which 
corrects each filter for (any) contaminating non-primary line and the continuum
and gives an image whose intensity units are photons cm$\rm^{-2}$ s$\rm^{-1}$ 
steradian $\rm^{-1}$.

\begin{figure}
\includegraphics{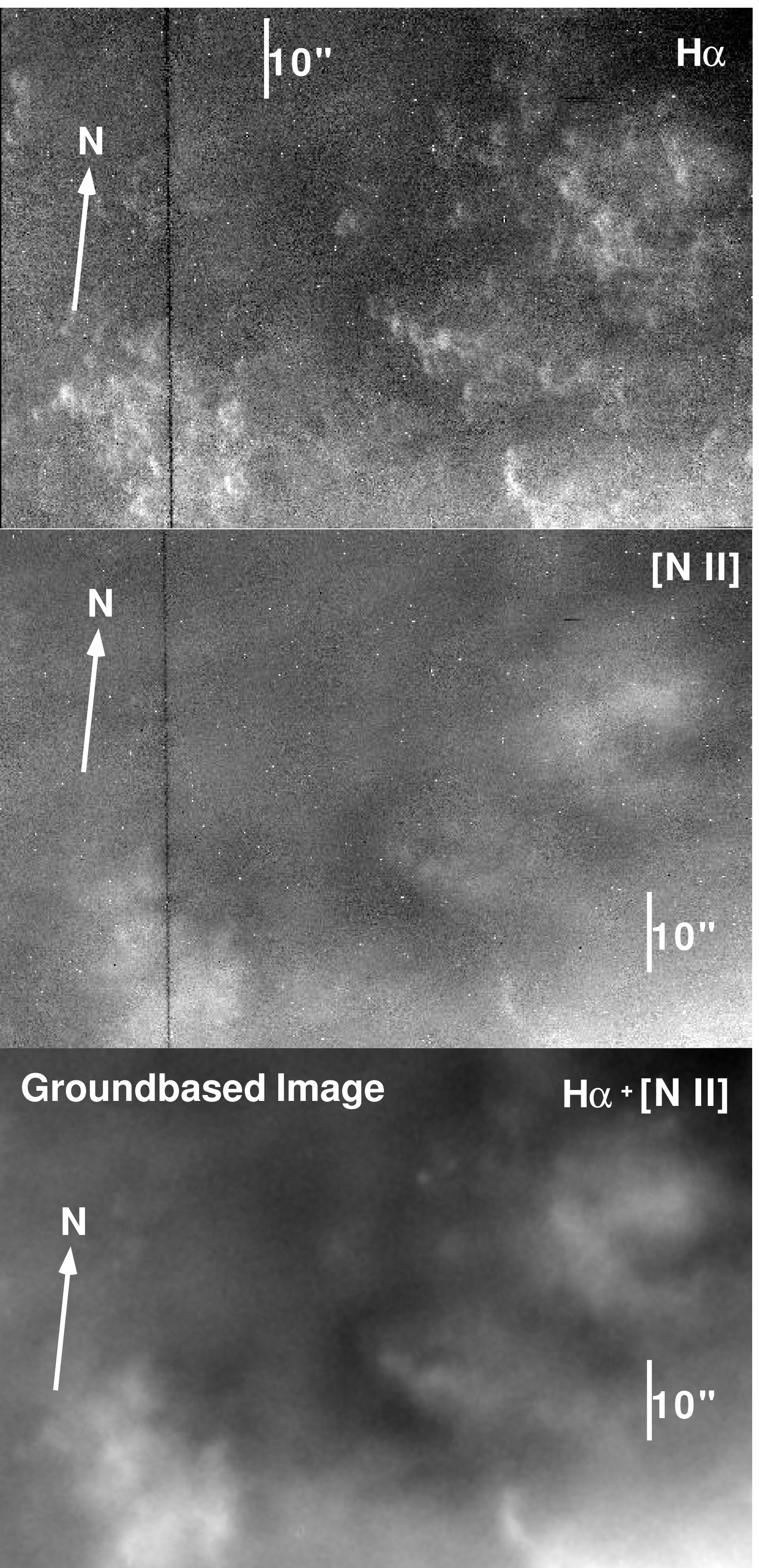}
\figcaption[2.2.1.A]{The top two images show a portion of the combined images in \Ha\ and [N~II]
created in programs GO-7286 and GO-9489. The field of view of each is 95\arcsec x66\arcsec\ and the vertical axis is pointed towards PA = 6\arcdeg. Contrary to most 
images of photoionized nebulae, the knots are much sharper in \Ha. The vertical dark
line is caused by the seam between WFPC2's CCD2 and CCD3. The lowest image is the same field
in an unpublished \Ha +[N~II] image made by O'Dell with the MOSAIC camera on the CTIO
4-meter telescope under seeing conditions of fwhm = 0.9\arcsec.
}
\end{figure}

The \Ha\ and [N~II] images are shown in Figure 3. 
There are many indications of knots of emission, although only a few well defined
bright cusps are seen. The features are much sharper in \Ha\ than in [N~II], an
unusual trend which is consistent with the result found in an earlier study using
both slitless spectroscopy and imaging (OHB2000).
The lowest portion of Figure 3 gives 
a comparison of the HST WFPC2 images with a groundbased \Ha +[N~II] image 
made with the MOSAIC camera of the 4-m Blanco Telescope of the Cerro Tololo 
Interamerican Observatory (CTIO). This image was made during a period
of astronomical seeing with a measured fwhm = 0.9\arcsec\ and a pixel size of
0.27\arcsec. 
We see that the bright regions of
about 15\arcsec\ diameter are broken up into multiple and often overlapping bright
cusps with chord sizes about 2\arcsec.
The image shows only diffuse faint emission in [O~III], which is not surprising
since it lies outside of the primary zone of [O~III] by the nebula and the knots
are known to emit little energy in this ion (M1992, OH1996, OHB2000).

\subsubsection{NICMOS Observations}

We also made parallel observations with all three cameras of the NICMOS instrument.
Because of the filter combinations available and the lack of simultaneous
focus, in part due to an error in the observing program, 
only the NIC3 detector results were useful for this study and even its images are out of focus. 
That camera has an array of 256x256 pixels, each subtending 0.2\arcsec x0.2\arcsec. We made four 
exposures (total exposure times of 10,752 seconds in each) in both the F212N filter and the F215N filter.
The F212N filter primarily isolates the \htwo\ 2.12 
\micron\ line and the F215N filter the adjacent continuum. The field imaged is shown
in Figure 1 and is almost identical with the field observed in program GO-7286.
These new images were not averaged with the older data because there is 
always a certain amount of degradation of resolution in combining
two not-identical-pointing images, the new images are better spatial resolution,
and the earlier observations were of only 3326 seconds (F212N) and 3582 seconds 
(F215N).  

The NIC3 images are significantly out-of-focus, but some information can be obtained
from them. 
The stars in our field of view have central dips of about
30\%\ and a full width half maximum (fwhm) value of 9 pixels. 
This means that we cannot hope to resolve nebular \htwo\ structures of less
than 1.8\arcsec. 

The images were subjected to the STScI's pipe-line data processing, which gave
corrected count-rates (CR) per pixel in each detector. The instrument data handbook
(section 5.3.4) gives for the calibrated flux (Flux, in ergs \pers\ \cms) from an emission line
Flux=1.054xFWHMxPHOTFLAMxCR, where FWHM is the full width at half maximum of the
filter and PHOTFLAM is a sensitivity constant determined from observations of 
a calibrated continuum source. In using this equation the assumption has been
made that the emission line is at the center of the filter (a good assumption for
this low velocity source) and that the continuum contribution has been subtracted
(discussed in this section).  Inserting the tabulated values of FWHM = 202 \AA\ and 
PHOTFLAM = 2.982 x 10$\rm^{-18}$ gives Flux = 6.35 x 10$\rm^{-16}$.  We corrected
for the continuum by examination of a section of the image that contained no 
evidence of knots, normalizing F215N filter signal to be the same as the F212N
image in that section, then subtracting the normalized F215N image from the F212N
image. After scaling into surface brightness (\Stwo\ in \sbunits)
and trimming a bad edge, one has the image shown in Figure 4, where the
average value of \Stwo\ is 1.0 x 10$\rm^{-5}$ \sbunits\ and the peak value is 5 x 10$\rm^{-5}$ \sbunits.
The intrinsic values of \Stwo\ will be higher by an amount determined by the defocus.
However, we should be able to determine
accurate total fluxes for the individual knots. These points are addressed in \S~4.3, where we
also compare our values with groundbased observations. 

\begin{figure}
\epsscale{1.0}
\includegraphics{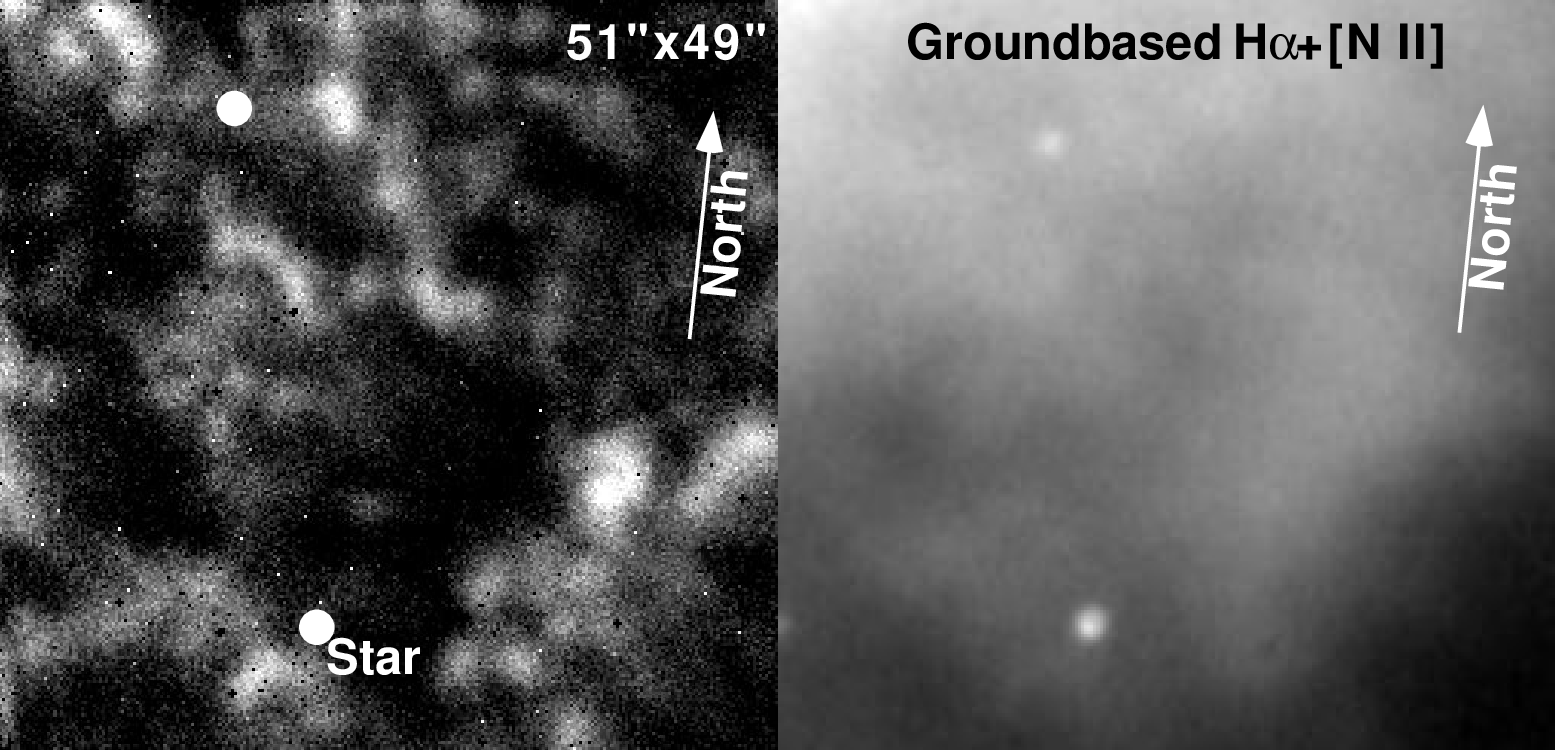}
\figcaption[2.2.2.A]{
The left image is the out-of-focus 51\arcsec x 49\arcsec\ NIC3 parallel image in continuum subtracted
values of \Stwo. The image has been rotated so that the vertical axis is at 
PA=6\arcdeg\ in order to nearly match the orientation shown in Figure 1. The
average value of \Stwo\ is 1.0 x 10$\rm^{-5}$ and the peak value is 4 x 10$\rm^{-5}$.
Correction of the peak values of \Stwo\ for the instrumental fwhm being greater than the
cusp shaped area of the emitting regions is discussed in the text.
The characteristic fwhm of the stars  is about 1.8\arcsec. The right image is the matching field
from the same \Ha +[N~II] image used in Figure 3.}

\end{figure}



\section{ANALYSIS}

\subsection{Analysis of the WFPC2 Images}

Studies of the variation in the peak surface brightness of the cusps with angular distance
from the central star have been made previously (OH1996, L\'opez-Mart\'{\i}n  \etal\ 2001) 
within the expectation that this would be a useful diagnostic of the knots.
This is because each cusp represents a local ionization front and
to first order their surface brightness in a recombination line
such as \Ha{} should scale linearly with the incident flux of LyC
photons so long as the cusps are in static ionization
equilibrium, with recombinations balancing photoionizations.
OH1996 demonstrated that the cusps were significantly fainter
than expected from this simple argument and L\'opez-Mart\'{\i}n
\etal\ (2001) found from the careful study of a few knots that
this deficit of emission could be explained by advection of
neutral hydrogen from the core into the cusp, which represents an
additional sink of ionizing photons and leads to a reduced rate
of recombinations compared to the static case.

In the light of the fact that we now have calibrated \Ha\ and [N~II] data that cover a much wider 
range of angular distances from the central star than in other studies,
we have re-executed an analysis of the brightness of the
cusps.

We have measured the peak cusp surface brightness for 530 knots identified in the WFPC2
fields of programs GTO-5086, GO-5311, and GO-5977 (OH1996, OB1997) and an additional 70 
in the WFPC2 field obtained in this study (\S~2.2.1).  The brightness of each cusp was 
measured in a 2x5 pixel box centered on the brightest portion and oriented with the long
axis perpendicular to a line towards the central star.  The nebular surface brightness 
used for correction of the cusp signal was determined from a nearby 5x10 pixel box. A 
similar analysis of the [N~II] image was done for each object, but the cusp sample box was
shifted 0.8 pixels towards the central star. The knot cores were also sampled in [O~III]
using 5x5 pixel boxes that were shifted 7 pixels away from the central star with respect to the
\Ha\ cusp. 

\begin{figure}
\includegraphics{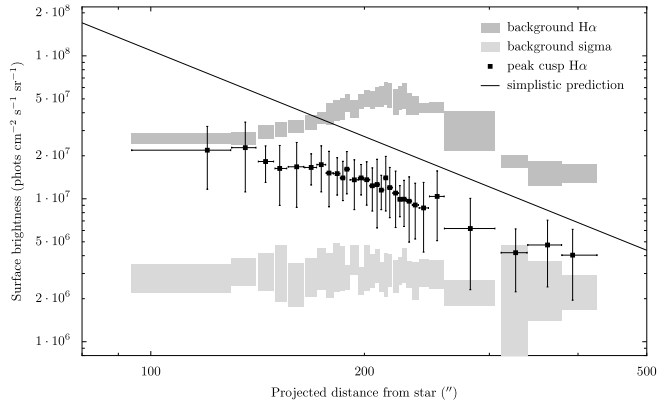}
\figcaption[3.1.A]{
The peak \Ha\ surface brightness of the knot cusps (filled circles), the nebular brightness
(dark grey boxes), and the standard deviation of the nebular brightness 
(light grey boxes) are shown as a function of the 
distance from the central star. The boundaries of the boxes indicate the one sigma
boundaries about the mean values.
The solid line is the predicted relation under the 
assumptions of the knots being viewed face-on, the distance from the central star being that
indicated by the angular separation, and diminution of LyC radiation by advected neutral
hydrogen being unimportant. In the case of failure of each of these assumptions, the 
expected surface brightness would be lower than this prediction.}

\end{figure}

The results for \Ha\ of this analysis are shown in Figure 5. In this figure we show
the surface brightness for \Ha\ and the standard deviation of the nearby nebular emission
for a box of the same size as used for extracting the cusp brightness.
The results indicate that we have traced the cusps out to where they are lost in the 
fluctuations of the nebular background. The upper line is the expected relation if all of the
knots were viewed face-on and were located at the distance indicated by their angular 
separation (these are mutually exclusive conditions). This is the same method of calculation as OH1996, except the more recent value of
the total flux of the nebula of F(\Hb)=3.37x10$^{-10}$ ergs \cms\ \pers\ (O1998) was used. This method of calculation 
also ignores the effects of advection. In practice, few of the knots will be viewed in the plane of the sky.
Most will be at foreground or background positions that will place them further from the
central star than indicated by their angular separations, and some advection will be present.
Each of these factors will cause the expected surface brightness to be less than the 
line labeled ``simplistic predictions''. Limb brightening would tend to work in the opposite 
direction by raising the cusp brightness, but this is evidently less important than the first two
effects.

\begin{figure}
\includegraphics{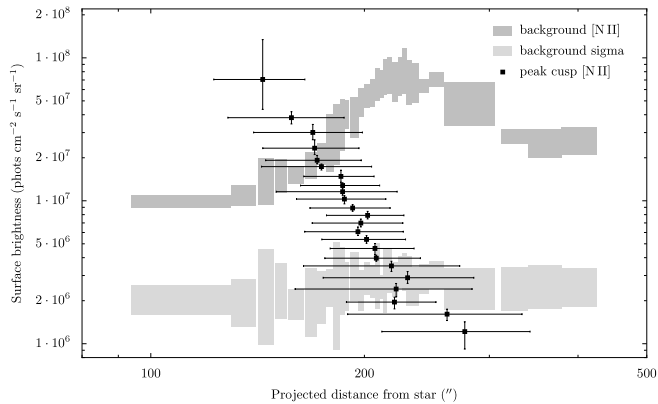}
\figcaption[3.1.B]{
This figure is similar to Figure 5, except that it presents the results of the analysis
of [N~II]. In addition to the [N~II] for the cusps (filled squares), we also give
the values for nearby portions of the nebula (filled boxes). 
The cusp [N~II] emission becomes weaker at
greater distances and the nebular emission shows a relative increase.
Negative values indicates that the [N~II] appears fainter than the nebular surroundings.}

\end{figure}

The characteristics of the nebula and the knots in [N~II] are summarized in Figure 6., where we see
that the cusp brightnesses monotonically decrease with increasing distance from the central star,
while the nebular brightness peaks at about 210\arcsec. The nebula's peak is due to reaching the 
ionization front that confines the inner-disk and outer-ring, these being seen at about the same 
distance from the central star in the northern sector (OMM2004), which numerically dominates this sample.

It is difficult to determine a trend of [O~III] emission from in front of the bright cusps,
but, there is a weak correlation of increasing emission with decreasing projected distance
from the central star.

\begin{figure}
\includegraphics{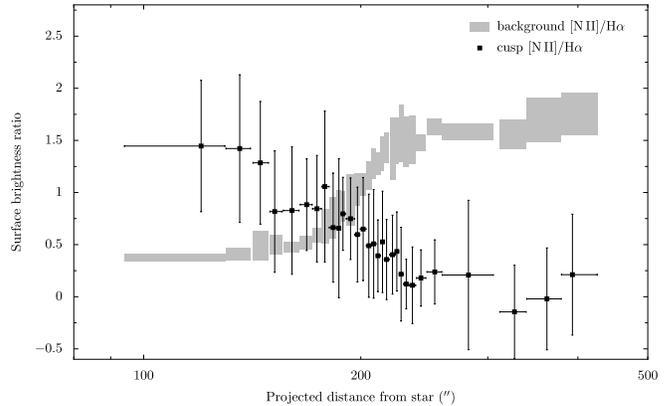}
\figcaption[3.1.C]{
This figure gives the 
[N~II]/\Ha\ ratio at the position of the knots.}

\end{figure}

The ratio of [N~II] lines and the \Ha\ line cusp emission is shown in Figure 7. 
We see that [N~II] becomes fainter relative to \Ha\ with increasing distance, whereas
in the nebula the opposite is true.
Each cusp is a microcosm of the nebula as a whole and should show
all the ionization stages up to the highest stage seen in the
local surrounding nebula. However, although the large-scale
nebula is probably very close to photoionization equlibrium, the
cusps themselves most certainly are not if they do indeed
represent photoevaporation flows. This is because the dynamical
time for the gas to flow away from the cusps is the same order as
the photoionization timescale. 

For gas in static photoionization equilibrium the ionization
state at a given point is governed by the balance between
recombinations and photoionizations, according to the local
electron density and radiation field, with the latter being
strongly affected by the photoelectric absorption of H and
He. For example, at larger radii in our observations the line of
sight progressively becomes dominated by the helium neutral zone
(from which [N~II] emission primarily arises) as the ionization
front of the inner-disk and the outer-ring is approached.

On the other hand, for very strongly advective flows, such as are
found in the knot cusps, recombinations are unimportant and the
ionization state at a given point is given merely by the
photoionization rate (cross section times ionizing flux) and the
length of time since the gas was first exposed to ionizing
radiation. Furthermore, the effective extreme ultraviolet optical depth to the
ionization front is low (of order unity) so that radiative
transfer effects are relatively unimportant, except for at low
ionization fractions. The simplicity thus gained, however, is
offset by the complication that the ionization state is now
intricately linked to the gas dynamics. One will still find an
ionization stratification but, since the cause is now different,
one may see a difference in the detailed distribution of the
ions.

The thermal balance in the ionized gas is also affected by the
advection since ``adiabatic'' cooling due to the gas acceleration
and expansion can become comparable to the atomic cooling.  The
relative importance of advection in the cusps is expected to
increase as the ionizing flux decreases so knots that are
farther away from the star might show lower temperatures in their
photoevaporation flows. This is one possible explanation for the
reduction in the [N~II]/H$\alpha$ ratio with distance,
which is explored more fully in \S~4.2 below. 

Knots in PN are almost unique in their exemplification of this
extreme regime of photoevaporation flows (Henney 2001). HII
regions are generally in the opposite ``recombination-dominated''
regime where the effects of advection are considerably subtler
(Henney \etal\ 2005b). The Helix knots are therefore an
important laboratory for testing our understanding of the physics
of this regime.

\subsection{Analysis of the STIS Slitless Spectrum Images}

As described in \S~2.1, the slitless images were made at three grating settings. In
several cases these included line doublets ([O~I] 6300+6364 \AA, [N~II] 6583+6548 \AA,
and [O~III] 5007+4959 \AA). Since each of these doublets arise from the same upper levels,
the ratio of intensity of the components is constant and in each case is close to three to
one. This means that we were able to derive higher quality images characteristic of each
ion by combining the separate images of lines, a possibility exploited further through the 
fact that the [O~I] lines appear at two grating settings,  with the result that images of 
four lines are available. The \Ha\ + [N~II] image is similar to that shown in Figure 1 
of OHB2000 in that the shorter wavelength [N~II] line image slightly overlaps \Ha.
We circumvented this problem by scaling, shifting and subtracting the image of the clearly
separated longer [N~II] line. The results are shown in our Figure 8. 
We see that the [N~II] emission is predominantly outside of the reference circle centered
on the knot, whereas the [O~I] emission peaks at roughly the same position as \Ha. The radii
and FWHM of the peak of the cusp emission are: [O~I] 0.52\arcsec\ and 0.21\arcsec, [N~II]
0.58\arcsec\ and 0.30\arcsec, \Ha\ 0.54\arcsec\ and 0.20\arcsec.

\begin{figure}
\includegraphics{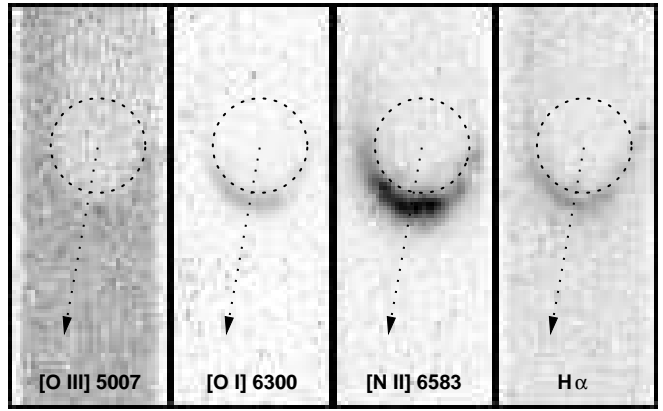}
\figcaption[3.2.A]{
Each of these panels shows the composite image for the various key ions in our analysis.
They were formed from adding the appropriately scaled images of forbidden line doublets and
the \Ha\ image was corrected for the slightly overlapping [N~II] 6548 \AA\ line's image.
The circle identifies the knot core and is in the same location in each panel. The arrow
indicates the direction towards the central star.}

\end{figure}

\begin{figure}
\epsscale{0.75}
\includegraphics{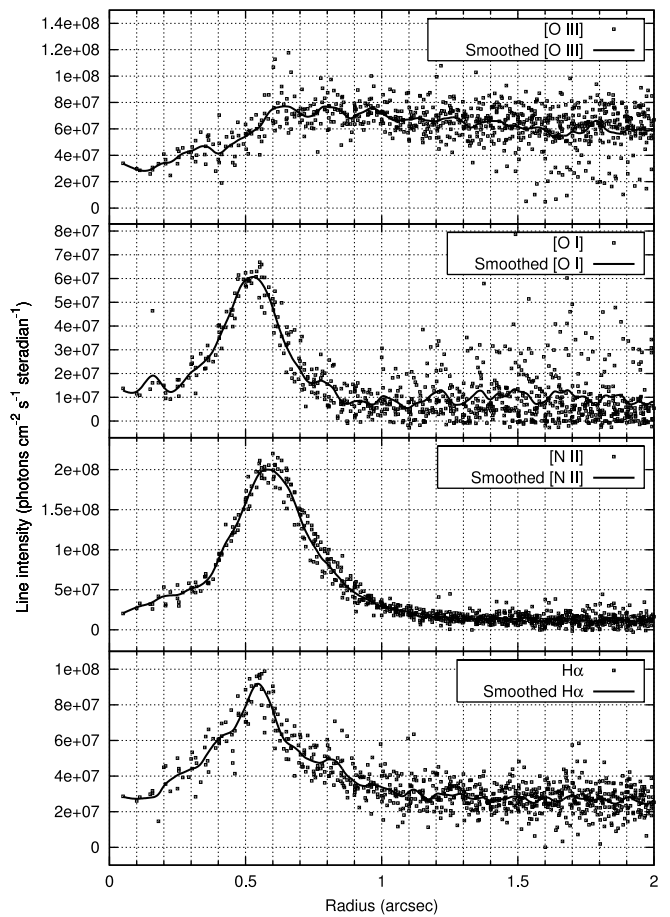}
\figcaption[3.2.B]{
Radial profiles of emission line surface brightness images of knot 378-801 as measured from the
nominal knot center. Symbols show all pixels whose radius from the knot center makes an angle
$\theta$
of less than 30\arcdeg\ with respect to the direction to the central star.  A heavy solid line
shows a smoothed version of the same data, while the lighter lines show 15\arcdeg\ subranges
in $\theta$  between 0\arcdeg\ and 90\arcdeg.}

\end{figure}

A more thorough presentation of the characteristics of the cusp are shown in Figure 9.
In this figure we show the profiles of each of the ions, now including [O~III], for all
of the data within 30\arcdeg\ from the direction to the central star.  The different
cusp widths and location of their peaks is well illustrated. It is obvious that the 
progression is not that expected from a simple ionization front, where the [O~I] emission
would be much narrower and clearly displaced towards the knot center, the [N~II] emission
would be further out and only slightly wider, and the \Ha\ emission would be broadest and
peak furthest from the knot center.

The explanation for this anomalous progression probably lies in
the strong departures from static photoionization and thermal
equilibrium in the cusps, as discussed in \S~3.1. OHB2000 could
successfully reproduce the distribution seen in earlier
observations by assuming a gradual rise in temperature of the gas
as it becomes ionized. This is in contradiction to the usual
structure of an ionization front, in which the gas temperature
rises to close to its equilibrium ionized value ($\sim 10^4$~K)
\emph{before} the gas becomes significantly ionized (Henney
et~al.\ 2005b). The OHB2000 model is unsatisfactory in that the
temperature structure was imposed in a totally ad~hoc
manner. This is rectified in \S~4.2 below, where we present
self-consistent radiation hydrodynamic models of the flows from
the cusps.

\subsection{Analysis of NIC3 \htwo\ Images}

The best resolution groundbased study of \htwo\ in the entire Helix Nebula is that of
S2002, who imaged the entire nebula. Although their angular resolution
is not stated, the pixel size employed was 2\arcsec\ and the astronomical seeing was probably no worse than that,  which means their effective resolution it is comparable to the
degraded focus fwhm of this study. Their image shows a few bright features within
our NIC3 field, but our images go fainter than their limit and show many more 
knotty structures. Except for the east-west feature in the upper central portion
of our Figure 4, there are no other features that correlate with the groundbased
image.

In our images there is no 
indication of the cusp structure seen in the optical emission lines,
which all arise from photoionized gas, but there are numerous emitting knots with
a characteristic fwhm of 2.3\arcsec. In no cases do we see the central dips 
characteristic of the defocused star images, which means that the intrinsic size of the 
\htwo\ cores are larger than a few pixels (about 0.5\arcsec). A Gaussian subtraction of
the stellar fwhm (1.8\arcsec) from the average knot fwhm (2.3\arcsec ) indicates an
\htwo\ emitting core of fwhm = 1.4\arcsec, which is comparable to the size of the cores
of the knots as outlined by extinction in [O~III]. We do not see an elongation of the
knots along a direction
perpendicular to a line directed towards the central star which is what we would
expect if this emission came from a cusp-shaped form. However, the theoretical expectation
is that the \htwo\ emission should come from a cusp shaped \htwo\ zone lying between
the emission line cusp and the central core of the knot. These cusps would have to
have elongations almost equal to the instrumental fwhm in order for us to have seen them.
When we do see evidence of 
cusp structures, they are always large, knotty, and have random orientations, indicating
that they are chance superpositions of individual \htwo\ emitting knots. 

An examination of F212N images of a southeast outer ring section of the Helix Nebula made as
part of program GO-9700 is useful. These images are in good focus, with stellar images
of fwhm=0.3\arcsec; however, the exposures showing knots were only 384 seconds duration
and a thorough discussion of those images will not be made here.
The \htwo\ emission appears as cusps of the same form as those seen in ionization in the
inner nebula. These cusps are about 0.4\arcsec\ thick and 1.8\arcsec\ wide. As noted in
the previous paragraph, this type of image would not produce any noticeable elongation of
our reduced resolution images.

The images show that the brightest knots in our NIC3 field have values
4 x 10$\rm^{-5}$ \sbunits, which is much less than the values for the same region
of greater than 10$\rm^{-4}$ \sbunits\ indicated in Figure 6 of
S2002. The instrumental broadening of both sets of data could mean that the
instrinsic surface brightness of the knots is higher than we give, but the comparable
spatial resolution of both means that this cannot be the explanation of the differences between them.
S2002 does not report that any correction was made for continuum radiation in their \htwo\
filter, which could indicate that the overestimate could be due to underlying 
continuum.  In our filter system the signal from the F212N filter was about
80 \%\ due to continuum, as determined from the F215N filter. S2002 did not detect
the Br $\gamma$ line at 2.166 \micron, with an upper limit of  
7 x 10$\rm^{-8}$ \sbunits, which means that it is not a source of contamination
of the F215N filter that we have used as a continuum reference. 

We can approximately correct our peak surface brightness values using the information from
the GO-9700 images. If the cusps have a rectangular size of 0.4\arcsec x 1.8\arcsec\ and our circular
images have a diameter of 2.3\arcsec, then the scaling factor from our observed peak surface
brightness to the intrinsic peak surface brightness will be $\pi$1.15$^{2}$/0.4x1.8=5.8.
This means that the observed peak surface brightness in our images (4 x 10$\rm^{-5}$ \sbunits)
would have intrinsic (corrected) values of 2.3 x 10$\rm^{-4}$ \sbunits.

A comparison of the HST NIC3 \htwo\ image with the CTIO  groundbased \Ha +[N~II] image 
is made in Figure 4. 
In this figure we see that there is no obvious correlation of the ionized gas
emission with the \htwo\ emission. This far out from the central star the radiation
is dominated by low ionization emission (O1998, Henry \etal\ 1999), which would be [N~II] in this
case, and the surface brightness of the ionized cusps will have dropped to well below
what can be resolved against the nebular background (for reference, see the discussion
in \S~4.1).  
   
\section{Discussion}

In this section we present and discuss the interpretation of our observations. In order to make the best use
of them, we have developed extensive theoretical models, the first combining radiation and hydrodynamic
effects, in support of explaining the optical emission line observations, and the second is a static 
model of the molecular region, in support of explaining the \htwo\ observations.  We combine our new
observations, our new models, and existing infra-red and radio observations to develop a comprehensive
model of the knots in the Helix Nebula.

\subsection{A General Model for 378-801}

\begin{figure}
\epsscale{0.75}
\includegraphics{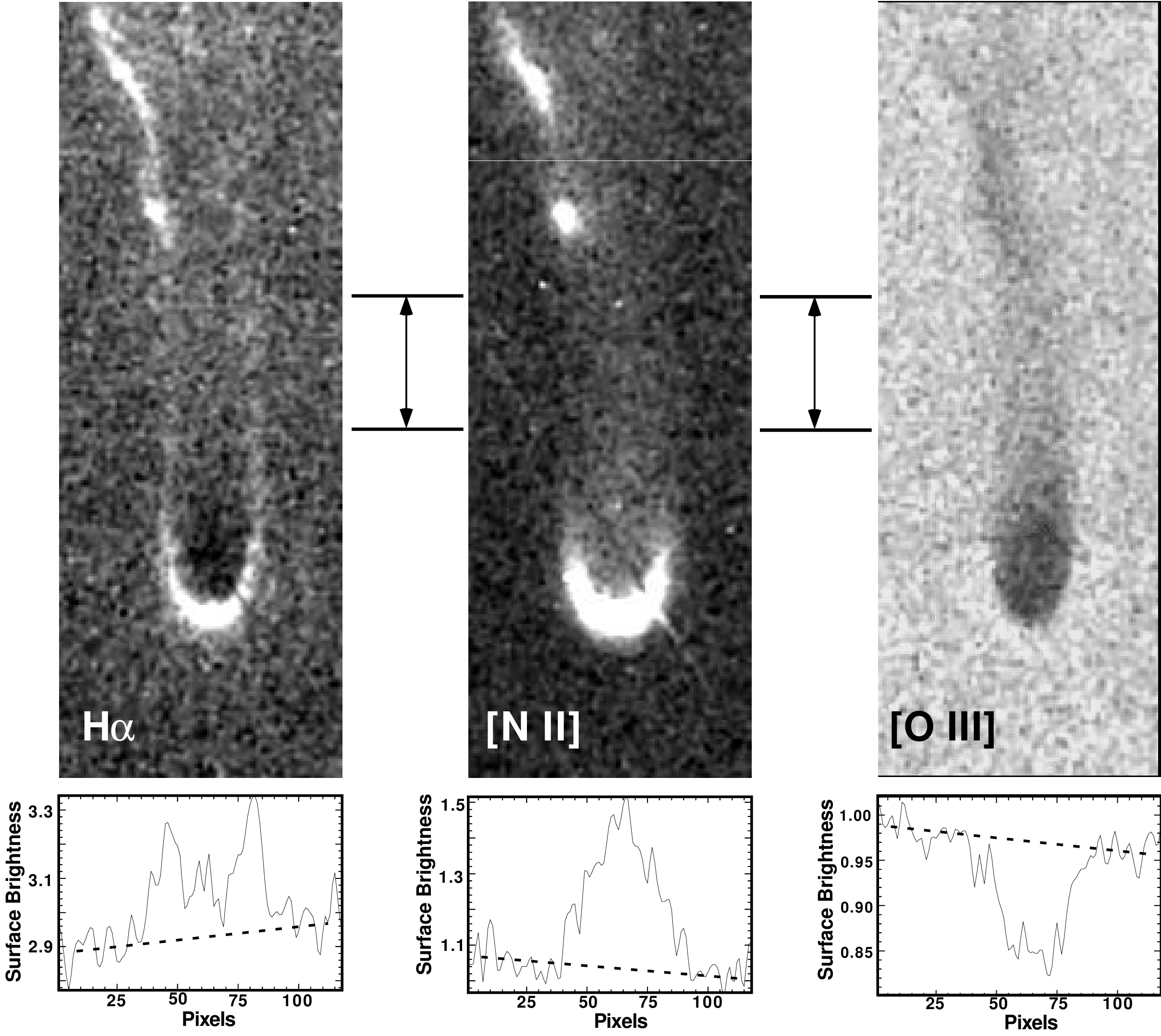}
\figcaption[4.1.A]{
The upper three panels show 5.9\arcsec x 16.2\arcsec\ samples around 378-801 from the WFPC2 programs GTO-5086 and GO-5311 
selected to lie along the symmetry axis of the tail, which coincides with a radial line drawn towards
the central star. The pixel scale has been adjusted to 0.05\arcsec\ in order to agree with that
of the STIS images.  The diagonal feature across the [N~II] cusp is an artifact in the original
image. The upper left feature in the \Ha\ and [N~II] images is probably a partially shadowed second knot that is unrelated to
378-801.  The bands between the upper panels show the range of rows that were averaged when
making the tail profiles shown in the lower three panels. The limb brightening seen in \Ha\ contrasts
with the concentrated emission seen in [N~II] and the concentrated extinction seen in [O~III].
}

\end{figure}

We probably know more about the object 378-801 than any other knot in the Helix Nebula
because of the wide variety of observations that have been made of it. The best optical
image is that of OH1996, where the knot is the central object in their Figure 3.
That source was used for preparing the monochromatic images shown in our Figure 10.
The bright cusp is well defined in \Ha\ and [N~II] with a central extinction core visible in
[O~III] emission.  The tail is well defined, with nearly parallel borders. At 8.5\arcsec\
away from the bright cusp there is a secondary feature that that starts on the east side of the
tail and extends away from the central star without the bright cusp-form characteristic of 
other knots found this close to the central star.  Its shape indicates that this is not a feature
of the tail of 378-801, rather, that it is a second knot that lies nearly along the same radial
line from the central star.

\subsubsection{The Core}
A molecular core in the knots was predicted by Dyson \etal\ (1989) before observational measurement
and since then has been detected with increasing spatial resolution.
378-801 has been imaged in the CO J=1--0 2.6 mm line by H2002 with an elliptical 
Gaussian beam of 7.9\arcsec x 3.8\arcsec\ having an orientation of the long axis of PA = 14\arcdeg ,
that is, at 18\arcdeg\ from the orientation axis of the object. In that image one sees two peaks
of CO emission, one on-axis 3\arcsec\ from the bright cusp and a second associated with the 
overlapping feature, with a peak at 8\arcsec\ from the bright cusp. Since they used the lower 
spatial resolution groundbased images of (M2002) for reference, they interpret this feature as part
of 358-801's tail, which it clearly is not. They also observed the object in the \htwo\ 2.12 \micron\
line under conditions of seeing of 1.2\arcsec\ and found a slightly broadened source just inside the 
curved optical bright cusp, having a surface brightness of about 10$^{-4}$ \sbunits\ (using the S2002
images for calibration). This image is consistent with the small cusps one sees in the southwest
region of the Helix imaged in \htwo\ as part of program GO-9700 (\S~3.3).
The CO source in the tail of 378-801 has a peak emission at \vsun 
=31 \kms\ (H2002 give \vsun =V$\rm _{LSR}$ - 2.9 \kms\ and they work in LSR velocities).  
This velocity agrees well with \vsun = 31.6 $\pm$1 \kms\ derived from optical spectra by M1998,
which means that there is not a large relative velocity of the CO source within the tail and its
associated knot.  

\begin{figure}
\epsscale{1.0}
\includegraphics[angle=90]{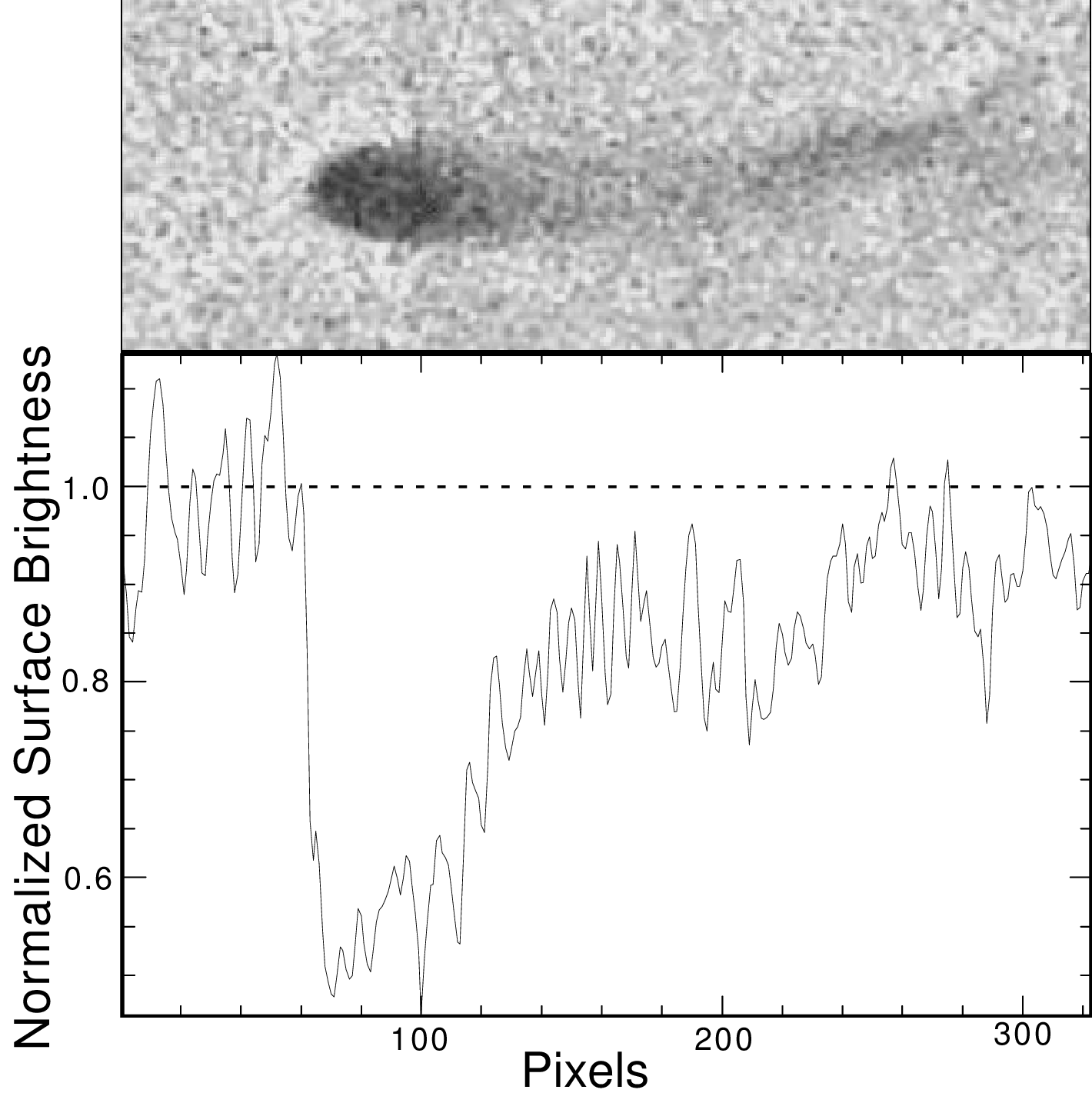}
\figcaption[4.1.1.A]{
The top panel in this figure is the same as the [O~III] image in Figure 10. The lower panel
shows a trace 0.15\arcsec\ wide along the symmetry axis of the cusp and tail. This illustrates
not only 
the concentration of extinction to the core of the knot, but that there is material that extends
out to where the the adjacent source confuses matters at about pixel number 200.
}

\end{figure}

The distribution of dust within 378-801 is best determined by tracking the extinction of background
nebular emission in the light of [O~III], since the knot has little intrinsic [O~III] emission.
Figure 11 shows a trace of the surface brightness along the symmetry axis of the knot and its
tail, after normalizing the local background to unity. The maximum extinction gives a contrast
of 0.5, which corresponds to an optical depth of 0.69. This value will be a lower limit if the
knot does not lie in front of all the nebular [O~III] emission. That seems to be the case, as
M1998 give high resolution spectroscopic evidence (their \S~5.1) that the knot is causing extinction
only in the redshifted component of the nebula's light.  This value of the optical depth is 
probably underestimated since the nebular light will slightly fill-in the central region due to
the finite point spread function of the WFPC2 (McCaughrean \&\ O'Dell 1996). Using the method
of determination adopted by OH1996 (which is similar to that of M1992) and a gas to dust mass
ratio of 150 (Sodroski \etal\ 1994), gives for a lower limit to the mass within the entire
core (a sample of 2.1\arcsec x 2.3\arcsec\ centered on the core) of 3.8x10$^{-5}$ M$\rm _{sun}$.
This is similar to the lower limit of 1x10$^{-5}$ M$\rm _{sun}$ found from the CO observations for 
378-801 and its nearby trailing feature (H2002). The peak extinction corresponds to a lower limit
of the column density of hydrogen of 1.7x10$^{21}$ hydrogen atoms \cms. If this material is
concentrated to a distance corresponding to 1\arcsec\, then the central hydrogen density of the
knot is at least 5.2x10$^{5}$ \cmq.  

As shown in \S~3.2, the appearance of \htwo\ emission in a cusp just behind the ionized cusp
is what is expected for a progression of conditions going from the outer ionized gas, through
atomic neutral hydrogen, then \htwo.  The innermost region of the core would be coldest and
have the most complex molecules.

\subsubsection{The Tail}

The appearance of the tail will be a combination of the properties of the gas and dust found
there and the photoionization conditions. A longitudinal  scan of the cusp and tail in [O~III]
was shown in Figure 11 and a cross sectional profile of a section of the tail was shown for
\Ha , [N~II], and [O~III] in Figure 10.  Taking the [O~III] extinction as a measure of the 
total column density and assuming axial symmetry, it appears that the
tail is a centrally compressed column of material that decreases slowly in density with increasing
distance from the center of the knot. The cause of this distribution is unknown, although it is consistent with the acceleration of the neutral gas via the rocket effect, e.g. (Mellema \etal\ 1998). The low relative
velocity of the core and the CO peak in the tail  (\S~5.1.1) argues that this reflects the
initial distribution of material following the formation of the core through an instability
followed by radiation sculpting of the tail (O2002)

Within the LyC shadow of the core the conditions will be very different from those in the nebula.
This condition has been theoretically modeled by Cant\'o \etal\ (1998) and those models have been compared with
observations of the knots in the Helix Nebula and the the proplyds in the Orion Nebula (O'Dell 2000).
Material in the shadow will only be illuminated by diffuse photons of recombining hydrogen and helium,
with the most important being hydrogen. Because most of these recombinations will produce photons
only slightly more energetic than the ionization threshold of hydrogen, whereas the stellar 
continuum is emitted mostly at higher energies, the average energy per photon given to the 
gas in the shadow will be much lower than in the nebula and the electron temperature will be 
about two-thirds that of the nebula (Cant\'o \etal\ 1998, Osterbrock 1989).  If the gas density within
the shadow is sufficiently high, then there will be an ionization progression much like when one
approaches the ionization front of an H~II region, with a neutral core in the middle. If the
gas density is low, then complete ionization of the shadow region will occur.

Our profiles in \Ha\ and [N~II] across the tail shown in Figure 10 are very different. In \Ha\ we
see strong limb-brightening, as if we are seeing a cylinder of emission edge-on.  In [N~II] we
see that the emission is very similar to that of the total column density of material, as 
measured by the [O~III] extinction.  The well defined \Ha\ boundary makes it appear that only the
outer part of the tail is photoionized. However, to create the collisionally excited [N~II]
lines, one needs both electrons of several electron-volt energy and singly ionized nitrogen.
It is obvious that a more sophisticated photoionization model is required and the paper by
Wood, \etal\ (2004) is a step in this direction, although it is not directly applicable to this knot
because they only consider a shadow formed in a region surrounded by only singly ionized helium.

\subsection{Models of the Knots that combine Hydrodynamics and Radiation}

In order to model the properties of the ionized flows from the
cometary knots, we have carried out a preliminary numerical study of
the dynamical evolution of a dense neutral condensation, subject to
the effects of stellar ionizing radiation. The simulations were
carried out by means of the radiation-hydrodynamics code described in
Henney \etal\ (2005a). The initial conditions for the simulation were
a cylindrical\footnote{Although the initial shape of the globule may
affect its subsequent evolution and eventual destruction, the
properties of the ionized flow from the globule head should be
insensitive to this.} concentration of dense neutral gas, with core
radius $2\times 10^{15}$~cm and density $3\times 10^4$~cm$^{-3}$,
which was illuminated by the ionizing spectrum similar to that of the
Helix central star ($T_\mathrm{eff} = 120,000$~K; ionizing luminosity
$Q_\mathrm{H} = 7.8\times 10^{45}$~s$^{-1}$). The local diffuse field
is treated in the on-the-spot approximation and the global nebular
diffuse field, with an expected strength of only 1--4\% of the direct
radiation (Lopez-Mart\'\i{}n et al.\@ 2001), is neglected. The only
source of opacity considered in the simulations is photoelectric
absorption (the dust optical depth through the ionized flow ion the
ionizing ultraviolet is only of order 0.01 and can be safely
neglected). Free-expansion conditions were used on the grid
boundaries, which is reasonable since the ionized flow becomes
supersonic. A simple ram-pressure balance argument indicates that the
stand-off shock between the ionized globule flow and the ambient
nebular gas should occur at about 50 times the globule radius, which
is outside our computational grid. Any wind that may exist from the
central star is confined very close to the center of the nebula
(Zhang, Leahy, \& Kwok 1993) and does not affect the knots.

\begin{figure}
\includegraphics{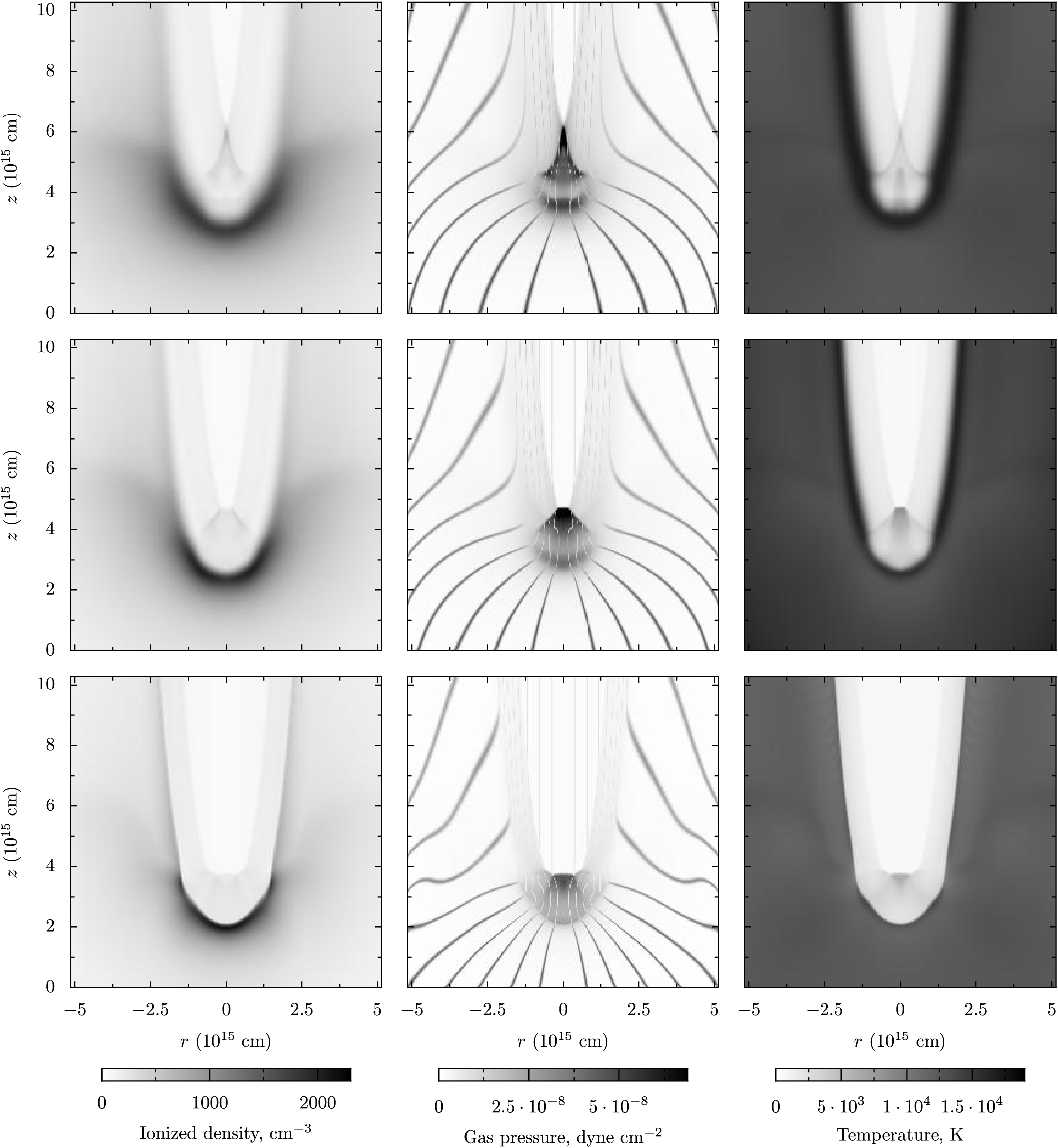}
\figcaption[4.2.A]{Physical structure of the model simulations at an age of
132~years for a knot at a distance of $4.63\times 10^{17}$~cm
from the ionizing star. The three rows of panels show simulations
calculated using different degrees of hardening for the ionizing
radiation (most hardening in the top row, no hardening in the
bottom row). The three columns show, ionized gas density, gas
pressure, and gas temperature, from left to right. The central
column also shows contours of the initial cylindrical radius of
the gas at each point, which serve as approximate streamlines for
the ionized part of the flow.}

\end{figure}

Models were run with the globule located at various distances from the
star between $3\times 10^{17}$ and $10^{18}$~cm, and the evolution was
followed for approximately 500~years. A snapshot of the structure of a
typical model is shown in Figure 12.

The photoevaporation of the surface layers drives a shockwave through
the knot, which compresses it and accelerates it away from the
ionizing star (Bertoldi 1989, Mellema \etal\ 1998, Lim \&\ Mellema 2003).
In our simulations, the neutral clump reaches
speeds of order 2 to 5 km~s$^{-1}$. At the same time, an ionized
photoevaporation flow develops from the head of the clump, which
accelerates back towards the star, eventually reaching speeds of order
20 km~s$^{-1}$. The bulk of the optical line emission is produced by
denser, slower-moving gas from the base of the photoevaporation flow,
near the ionization front, as can be seen in the left-hand panels of
Figure 12, which show the ionized gas density. Three
models are shown, which differ in their treatment of the hardening of
the radiation field, which is treated only very approximately in the
current simulations.  In the model with maximum hardening (upper
panels), the ionization front is very broad and the gas temperature
(right panels) has a pronounced maximum on the neutral side of the
front. When the hardening is reduced (lower panels), the front is much
sharper, with a less pronounced temperature peak. The pressure of the
photoionized flow (central column of panels) is also higher in the
models with greater hardening and, as a result, a stronger shock is
driven into the neutral knot. Weaker shocks are also driven in from
the flow from the sides of the knot, which eventually converge and
bounce off the symmetry axis, as in the upper model of
Figure 12.

Although these two-dimensional simulations do not contain all the
details of atomic physics that have been included in one-dimensional
models (e.g., Henney et al. 2005b), they nonetheless capture the
most important physics of the photoionized flow, which is dominated by
photoelectric heating and expansion cooling, and therefore they can
provide a tool for investigating the physical basis of the observed
optical emission properties of the knots. The most important of these
are that the [N~II]/H$\alpha$ ratio falls precipitously with distance
of the knot from the central star (\S~3.1) and that the [N~II]
emission from the best-studied individual cusps is found to lie
slightly outside the H$\alpha$ emission (\S~3.2).  OHB2000 attempted
to explain the second of these by positing an ad~hoc broad temperature
gradient in the photoevaporation flow, leading to enhanced collisional
line emissivity and depressed recombination line emissivity at greater
distances from the knot center.

We have calculated for each model the emission properties of the cusps
as follows. First, we automatically identified the ridge in the
ionized density at the cusp and fit for its curvature. Using the
position and curvature of the cusp we determined a nominal center for
the knot.  Then, we derived radial emissivity profiles as a function
of distance from the knot center, averaging over all radii within
60\arcdeg\ of the symmetry axis. From these profiles, we calculated
the mean surface brightness of the cusp and its mean radius from the
knot center for each emission line. 
These steps were carried out for each of a sequence of times in the
evolution of the knot up to an age of $\simeq 500$~years, which were
finally averaged to give a global mean and standard deviation for each
model.\footnote{In order to explore the mechanism of formation and
evolution of the knots, it would be desirable to perform fuller
simulations that tracked the knots over thousands of years. However,
our present simulations have only the more limited goal of
explaining the properties of the ionized flow. For this purpose, we
feel that the baseline of 500 years is sufficient, which represents
about 15 dynamic times for the ionized flow.}  In the final
averaging the initial 50~years of evolution were omitted since this
represents a highly non-steady phase in which the flow is still
adjusting to its quasi-stationary configuration.

The lines that we considered were H$\alpha$ and [N~II]
6583~\AA, with temperature-dependent emission coefficients that were
calibrated using the Cloudy plasma code Ferland (2000).
The ionization fraction of nitrogen was assumed to exactly follow that
of hydrogen and double-ionization of nitrogen was neglected (the
extreme weakness or absence of [O~III] emission from the cusps
indicates that this is a reasonable approxmation). 

\begin{figure}
\epsscale{0.75}
\includegraphics{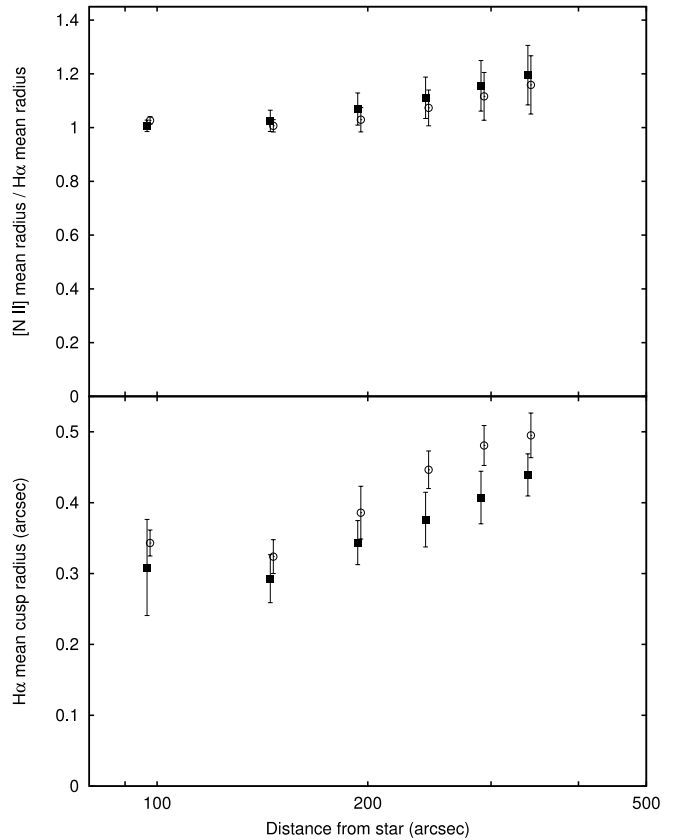}
\epsscale{1.0}
\figcaption[4.2.B]{Model predictions for knot cusp radius as a function of
distance from the central star. Lower panel: mean radius of
H$\alpha$ emission from cusp in arcseconds. Upper panel: ratio of
mean cusp radius in [N~II] to that in H$\alpha$. Vertical
bars indicate the range of values during the evolution of the
models. Different symbol types correspond to models with differing
treatment of the radiation hardening and photoelectric heating
(see text).}

\end{figure}

\begin{figure}
\includegraphics{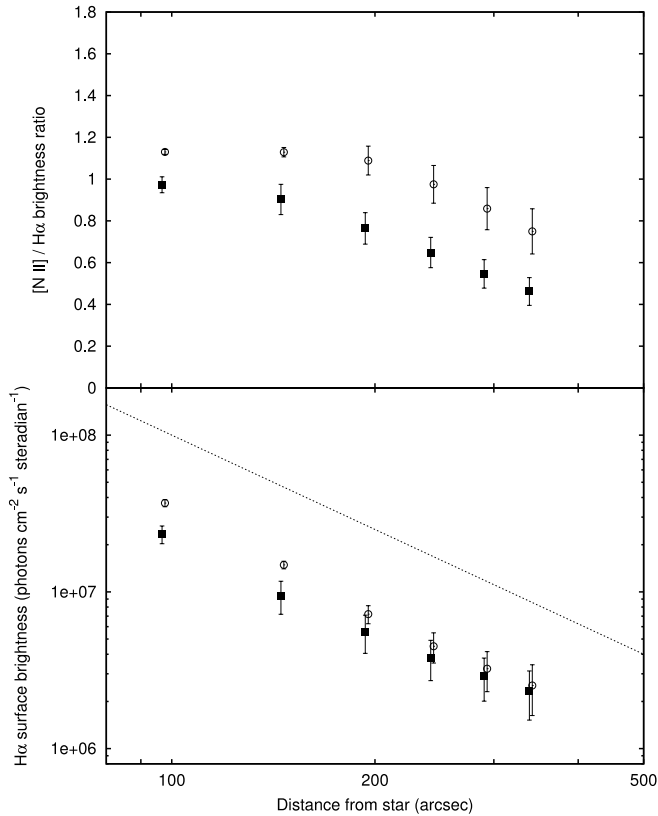}
\figcaption[4.2.C]{Model predictions for knot surface brightness as a
function of distance from the central star. Lower panel: face-on
surface brightness of H$\alpha$ emission from cusp. The dotted
line shows the simplistic prediction as in Figure 5. Upper
panel: ratio of [N~II] surface brightness to that of
H$\alpha$.  Symbol types are as in
Figure 13.}

\end{figure}

The results are shown in Figures 13 and 14 and for two different sets
of models, which differ in their treatment of the hardening of the
radiation field and in the contribution of helium to the photoelectric
heating. The upper panel of Figure 13 shows that the mean radius of
the [N~II] emission is indeed larger than that of H$\alpha$, but only
significantly so for knots that are farther than 200\arcsec{} from the
central star. This is a direct result of the temperature structure in
the ionized flow, which has a maximum close to the peak in the ionized
density. For the farther knots, this temperature peak is slightly
outside the ionized density peak, which biases the [N~II] emissivity
to larger radii. 

The upper panel of Figure 14 shows that the [N~II]/H$\alpha$ surface
brightness ratio of the knots does decline with distance, particularly
for the model with less hardening.  This is a direct result of the
flow not attaining such high temperatures when the knot is further
from the star, which is due to the increased relative importance of
``adiabatic'' cooling as the flow accelerates.  However, although this
behavior is in qualitative agreement with the observations, the
decrease seen in the models is much less sharp than is seen in the
real nebula. It should also be noted that we have had to assume a
nitrogen abundance of $\mathrm{N/H} = 1.2 \times 10^{-4}$ in order for
the absolute values of our [N~II]/H$\alpha$ ratio to be consistent
with the observations.  This is roughly two times lower than the
abundance that has previously been derived for the nebula as a whole
(Henry, \etal\ 1999). 

The N abundance that we find from our models is close to the solar
value and is very similar for the models with different hardening.
Although this could be interpreted as evidence for differing
abundances between the knots and the rest of the ejecta, we do not
think that such an inference is warranted.  The photoionization models
on which previous abundance studies have been based are only very
crude representations of the 3D structure of the nebula, so that their
derived abundances are probably not reliable.

The lower panel of Figure 14 shows the dependence on distance of the
H$\alpha$ cusp surface brightness of the models. The values shown are
for a face-on viewing angle and thus should by multiplied by a factor
of a few to account for limb-brightening. Once this is taken into
account, they are in very good agreement with the observed values
(Figure 5), and in particular fall well below the line derived from
equating recombinations in the flow to the incident ionizing flux,
which is an illustration of the importance of the advection of
neutrals through the front, as discussed in L\'opez-Mart\'{\i}n \etal\
(2001).

It is heartening that our simple radiation-hydrodynamic models are
qualitatively consistent with the trends seen in the observations,
although there are some quantitative discrepancies. In particular, in
the models, both the reduction in the knot [N~II]/H$\alpha$ surface
brightness ratio and the increase in the size of the [N~II] cusp
relative to the H$\alpha$ cusp only occur at large distances from the
central star, whereas in the Helix they are observed to occur at
smaller distances. Although this may be in part a projection effect,
due to the projected distances of the knots being smaller than their
true distances, that is unlikely to be the whole explanation. One
possibility is that the farther knots see a much reduced ionizing flux
due to recombinations in the nebula between them and the ionizing
star, whereas in our models we assume only a geometric dilution of
$1/r^2$. 

It is also possible that some of the atomic physics processes
that have been neglected in our models may prove important in
determining the thermal structure. More realistic simulations are in
preparation and it remains to be seen if they will lead to an
improvement in the agreement with observations.

\subsection{\htwo\ Emission}

In addition to the surface brightness in the \htwo\ 2.121~$\mu$m line discussed in \S~3.3, 
there is the important and constraining result of Cox et al. (1998), who determined from 
ISO spectrophotometry of six \htwo\ lines that this gas has an excitation temperature of
900 K. In this section we will show that radiation-only-heating models cannot explain such a high temperature.

The other basic constaint is the column density of \htwo.
If the molecular gas is in LTE then the surface brightness in an optically thin line is simply related to the gas column density. We adopt a 2.121~$\mu$m \htwo\ line surface
brightness of 2.3x10$^{-4}$ ergs \cms\ \psec\ sr$^{-1}$. At 900 K, we expect
\[
S(\mathrm{H_{2};\ 2.121~\mu m}) = 3.83 \times 10^{-24}
N(\mathrm{H_{2}})~\mathrm{erg\ cm^{-2}\ s^{-1}\ sr^{-1}}
\]
and obtain a column density of N(\htwo)$\simeq$ 6x10$^{19}$ \cms, or N(H)$\simeq$ 12x10$^{19}$ \cms\ if the gas is fully molecular.
This warm \htwo\ layer has a physical thickness of 1.2x10$^{15}$ cm (\S~3.2), so we find
a density of $\rm n(\mathrm{H_{2}})=N(\mathrm{H_{2}})/\delta r \simeq 6 \times
10^{4}~\mathrm{cm^{-3}}$. This is not significantly smaller than the density Cox et al. (1998) quote for \htwo\ to be in LTE. We assume this density in the remaining work.  

Since the emissivity is driven by the excitation temperature, the cause of the high temperature must be resolved
prior to comparing the observed and predicted surface brightness of the \htwo\ 2.121~$\mu$m 
line. We present our model in \S~4.3.1, give an interpretation of the observations and
derive the additional heating required in \S~4.3.2, and other relevant observations of the core of the knots are discussed in \S~4.3.3. 
A critique of previous comparisons of \htwo\ observations and models is presented in 
\S~4.3.4.

\subsubsection{The Predicted Knot \htwo\ Properties}

For simplicity, we assume that the knot is illuminated by the full radiation field of the central
star, a good assumption since the nebula is quite optically thin up to where knots are first detected.
We assume a seperation from the central star of 0.137 pc, appropriate for the knot 378-801. We
further assume that the knot has the same gas-phase abundances as the H~II region (Henry \etal\ 1999).
The illumination is assumed to be from a black body of 120,000 K and 120 $\rm L_\odot$ (Bohlin \etal\ 
1982, adjusted to a distance of 213 pc). In addition, we assume an ISM dust-to-gas ratio and grain
size distribution, but do not include PAH's since Cox \etal\ (1999) report that no PAH emission is seen.
We assume that the density is constant across the knot in our photodissociation calculations.
Cox \etal\ (1999) convincingly argue that the lower levels of \htwo\ are in LTE and that this requires
a density $n \ge 10^{5}$ \cmq, which is much higher than the characteristic (1200 \cmq) density in
the ionized zone (OB1997).

\begin{figure}
\includegraphics{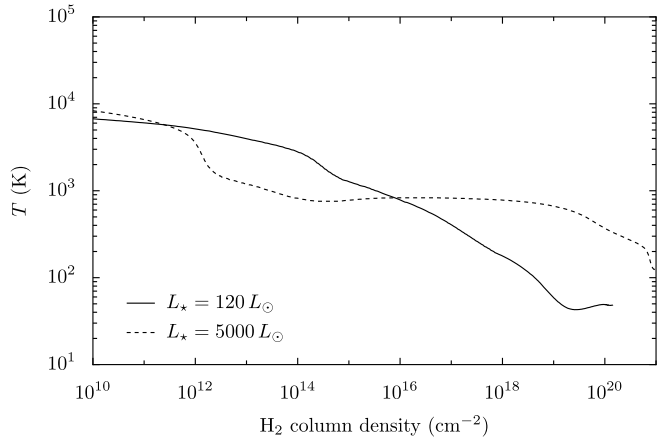}
\figcaption[4.3.1] {The gas temperature is shown as a function of molecular hydrogen column density
for our Cloudy models of the knot PDR. The solid line shows a model calculated using our best
estimates of parameters appropriate for the Helix knots. The dashed line is the t$=$4000~yr case 
considered by Natta \&\ Hollenbach (1998). The largest difference is that the central star luminosity
is about a factor of 40 larger. Grain photoelectric heating of neutral gas creates a significant
warm (about 900~K) region.}

\end{figure}

With these assumptions the conditions within the knot can be calculated
with no unconstrained free parameters. The solid line in Figure 15 shows the gas temperature as a 
function of the column density of molecular hydrogen as one goes into the knot. It can be seen that
the column of warm \htwo\ is very low (about 10$^{16}$ \cms\ for T$>$900 K) and that the temperature 
has fallen to about 50~K before an appreciable column density is reached, much lower than the 
observed \htwo\ temperatures.

This result seems to conflict with NH1998, who found significant amounts of warm (900~K) molecular 
gas in their model for a PN envelope at an evolutionary age of 4000 years.  We recomputed this case 
using the time-steady approximation and approximating the dynamics with constant gas pressure. 
While the time-steady assumption is not really appropriate for the \htwo\ transition region because 
of the long formation timescales, it is valid for both the atomic and fully molecular regions, and
serves to illustrate the important physics.  The dashed line in Figure 15 shows this calculation. 
While we do not reproduce their large region of hot (10$^{4}$~K) predominantly atomic H, we see the
essential features of the transtion and fully molecular regions. In particular, a significant amount
of gas has a temperature of about 900~K. This is due to grain photoelectric heating of the gas, as
NH1998 point out. Although this process does occur in our fiducial Helix model, it is not able to
sustain warm temperatures mainly due to the much lower luminosity of the Helix central star. The
cooling time in the molecular gas is very short (of the order of decades), so that time-dependent 
effects are unlikely to change this result. Indeed, NH1998 find a similar behavior at later 
evolutionary times, after the luminosity of the central star has declined.  For example, their model
results at an age of 7000~years no longer show a significant column of warm molecular gas.

In conclusion, it would be possible to achieve the derived column density of heated molecular hydrogen (6x10$^{19}$)
only by increasing the luminosity of the central star by more than an order of magnitude from the 
observed value. In the following section, we investigate the possibility that the warm \htwo\ may
instead be heated by shocks.

\subsubsection{Are Shocks the source of the Warm \htwo\ ?}

In this section we derive the rate of extra heating that would be necessary to explain the 900 K temperature
and assess whether this can be explained by shock heating.

\begin{figure}
\includegraphics{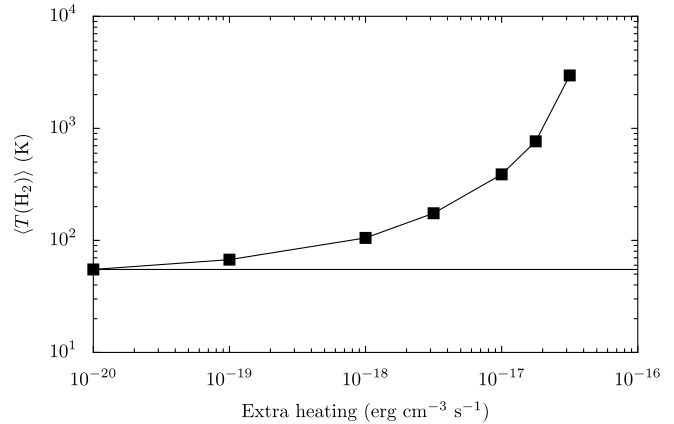}
\figcaption[4.3.2]{The enhancement of the temperature of the gas in our model knots is shown as a function of the added extra heating. The horizontal line is the expected value with no extra heating.}

\end{figure}

As shown in \S~4.3.1, the expected \htwo\ temperature is about 50 K for the conditions present in the Helix when only radiation from the central star and the x-ray source heat the gas.
Something else, probably mechanical energy, must add heat to the molecular gas to obtain
a temperature as high as 900 K. We quantify this by adding increasing amounts of extra heating
to our fiducial model. The results are shown in Figure 16. The extra heating is shown along the x-axis and the y-axis gives the \htwo\ weighted mean temperature. An extra heating rate of about 10$^{-16.8}$ erg \cmq\ \psec\  is necessary to account for the observed \htwo\ temperature. For comparison, the radiative heating across the \htwo\ region is 10$^{-19}$ erg \cmq\ \psec. Since the warm \htwo\ extends over a 
physical thickness of 1.2x10$^{15}$ cm, the extra power entering the layer to sustain the 
heating over this thickness is 1.9x10$^{-2}$ erg \cms\ \psec.

One source of non-radiative heating that naturally springs to mind is shock heating.
For the shock hypothesis to be viable, three requirements must be
satisfied:
\begin{enumerate}
\item The shock velocity should be just high enough to heat the gas up
  to $900$~K and excite molecular hydrogen emission.
\item The rate of energy dissipated in the shock must be sufficient to
  provide a heating rate of $2 \times 10^{-17} \mathrm{\ erg\ cm^{-3}\
    s^{-1}}$ in a layer of thickness $10^{15}\mathrm{\ cm}$.
\item In the case of a transient shock, the cooling time in the
  molecular gas must be sufficiently long that the gas remain hot for
  a significant time after the shock has passed.  
\end{enumerate}

For fully molecular conditions ($\gamma = 7/5$, $\mu \simeq 2.36$),
the first requirement is satisfied for shock velocities in the range
$4.5$--$5 \mathrm{\ km\ s^{-1}}$. The second requirement demands an
energy flux through the shock of $0.02 \mathrm{\ erg\ cm^{-2}\
  s^{-1}}$, which, combined with the first requirement, implies a
pre-shock hydrogen nucleon density of $\simeq 4.2 \times 10^4
\mathrm{\ cm^{-3}}$, and a post-shock density of $\simeq 2.6 \times
10^5 \mathrm{\ cm^{-3}}$. Within the margins of error, this density is
consistent with the values discussed above. Our hydrodynamic models of
\S~4.2 show that several shocks of this approximate speed are
indeed generated in the radiatively driven implosion of the knot.

However, the third requirement proves to be the most difficult to
satisfy, even in its weakest form, which requires that the time taken
for gas to pass through the thickness of the H$_2$ layer be less than
the cooling time, $ P / [(\gamma-1) L] \simeq 20$~years (where $L$ is the volumetric cooling rate). 
This implies that the particle flux through the layer be larger than the column
density divided by the cooling time: $ n v \gtrsim 2 \times 10^6
\mathrm{\ cm^{-3}\ km\ s^{-1}}$.  In our simulations, we find particle
fluxes that are at least ten times smaller than this value.
Furthermore, since the shock speed of $\simeq 5 \mathrm{\ km\ s^{-1}}$
exceeds the propagation speed of the ionization front at the knot's
cusp, the shocks quickly propagate up the tail and away from the cusp.
This is hard to reconcile with the observed location of the H$_2$
emission in 378-801 unless we are seeing this knot at a special time.

In summary, although shocks are initially attractive as a mechanism to
explain the observed H$_2$ emission, they seem to be ruled out by
cooling time arguments. The gas temperature is only enhanced in the
region immediately behind the shock, which is difficult to reconcile
with the observed location of the H$_2$ emission, which is immediately behind
the cusp ionization front. This objection could be avoided if the
shock were a stationary structure in the head of the knot, i.e., if
the shock and cusp were propagating at the same speed. However, such a solution
is very contrived and no such stationary shocks are ever seen
in our hydrodynamic simulations.\footnote{In the case of the Orion
proplyds, just such a stationary shock \emph{is} found, driven by
the back pressure of the ionization front acting on the neutral
photoevaporative wind from the accretion disk (Johnstone, \etal\ 1998.
However, the Helix knots do not possess
an ultra-dense reservoir of neutral gas such as is found in the
proplyds.}

\subsubsection{A Summary of Properties of the Core of the Knots}

H2002 have resolved the knot 378-801 in both \htwo\ and CO, finding that the \htwo\ emission occurs well displaced towards the bright optical cusp from the CO emission, which comes
from the core of the knot.  CO will only exist if \htwo\ is also present, since the 
chemistry that leads to CO is initiated by interactions involving \htwo. This means that the core emission must come from a cooler, higher density region than the \htwo\ emission. 
The optical depth for CO to be formed is about $\tau \simeq 4$ (Tielens, \etal\ 1993). If the optical depth is the factor determining the displacement of the peak of CO (about 3.5\arcsec $\simeq$ 10$^{16}$ cm), then the intervening column of gas has a density of about 10$^{6}$ \cmq. 

Dust extinction also provides an estimate of the total hydrogen column density through the center of the knot, presumably the core of the cool CO region. The observed extinction
($\tau \simeq $ 0.7, OB1997) corresponds to N(H)$\simeq$ 1.7x10$^{21}$ \cms, for an ISM dust-to-gas ratio. This dust extinction occurs across a region about 1\arcsec\ (3x10$^{15}$ cm),
so the density in the CO region is n(H)$\simeq$ 6x10$^{5}$ \cmq. The fact that the core extinction is less than that required to avoid photodissociation of CO argues that the core is
unresolved on our HST images and that the larger (10$^{6}$ \cmq) density applies.
This larger density is 
is an order of 
magnitude higher than our estimate of the density in the hot \htwo\ zone. Since the CO
temperature is probably about that of our model without the extra heating, the hot \htwo\
zone and the core are about in pressure equilibrium.

\subsubsection{Previous Comparisons of \Stwo\ with Evolutionary Models of Planetary Nebulae}

In a recent paper Speck \etal\ (2003, henceforth S2003) present new images of NGC 6720 (the Ring
Nebula) in \htwo\ 2.12 \micron\ at an unprecedented resolution of 0.65\arcsec.
They find that the Ring Nebula resembles the Helix Nebula in \htwo\ in that the
emission is concentrated into small knots, some of which were already known
to have tails (O2002), and these knots are arrayed in loops. It was already
known (O2002) that the Ring and Helix Nebulae have similar three dimensional 
structures and that the Ring Nebula is in an earlier stage of its
evolution, with much more extinction in the tails outside of the knots. This
last point argues that the tail material is residual material from the formation
process, rather than being expelled from the knot.

S2003 compare the average values of \Stwo\ for the Ring Nebula,
the Helix Nebula, and NGC 2346 with the predictions of an evolutionary model
for NGC 2346 derived by Vicini \etal\ (1999).  
In S2003's Figure 3 they compare the average values of \Stwo\ 
with the predictions of the Vicini models for various ages of the nebulae and 
argue for good agreement, even though the observed average surface brightness of the Helix Nebula is
 much larger than the predictions of the model, using their adopted age of 19,000 years.
If one uses the most recent determination of 6,600 years (OMM2004), then the agreement
becomes good for the Helix Nebula. However, if one uses their average surface brightness for NGC 6720 but the best value of
the age of 1,500 years (O2002), then that object is much too bright for its age.

We note that S2003's Table 1 gives an average value of \Stwo\ for
the Helix Nebula of 6 x 10$^{-5}$ \sbunits\ whereas the source they cite (S2002,\S~2.3)
gives an average value of 2 x 10$^{-4}$ \sbunits. The reason for this difference
is the the earlier, larger value refers to the average surface brightness of 
individual knots whereas the S2003 value refers to an average surface brightness over
an extended area (private communication with Angela K. Speck, 2005). The average brightness
of the knots in S2002 (2 x 10$^{-4}$ \sbunits) is comparable to the value of
2.3 x 10$^{-4}$ x 10$^{-4}$ \sbunits\ that we derived in \S\ 3.3 for the peak surface 
brightness of the knots.

There is reason to question acceptance of the procedure of comparison of 
the Vicini \etal\ (1999) model and the observed average surface brightnesses of the Ring and Helix Nebulae. The Vicini \etal\ model draws on the general theory for an evolving PN published by NH1998.
In both the Ring Nebula (S2003) and the Helix Nebula (S2002) one sees that \htwo\ emission
arises primarily from the knots, rather than an extended PDR behind the main
ionization front of the nebula. In their ``Summary and Conclusions'' section NH1998 point
out that knots do not adhere to their general model and would have a higher surface
brightness. However, since the knots cover but a fraction of the image of the 
nebulae, the average surface brightness is lower than the value for the individual
knots and depends on the angular filling factor. This means that one cannot compare
the average surface brightness of a knot-dominated nebula with the predictions of a
simple evolving nebula unless one has both a detailed model for the knots and an accurate
determination of their angular filling factor.

\subsection{On the association of the knots and CO sources}

There is reason to argue that all of the measured CO sources seen in the Y1999 
study are associated with knots. As the spatial resolution of the CO observations
have improved (Huggins \&\ Healy 1986, Healy \&\ Huggins 1990, Forveille \&\
Huggins 1991, Huggins \etal\ 1992, Y1999),
there has been a progressive ability to isolate individual
knots, with the best resolution study (H2002) targeted and found the target knot of this
study (378-801), as discussed in \S~4.3.3. Even at the lower resolution of the Y1999 study, spectra of individual
data samples commonly show multiple peaks of emission at various nearby velocities,
indicating that the unresolved regions are actually composed
of multiple emitters. A similar progression of improved resolution in \htwo\
observations has produced clear evidence for association with knots (S2002).
The isolation of individual knots becomes more difficult as one goes farther from the
central star because the numerical surface density of knots increases rapidly and 
the lower resolution CO studies appear amorphous first, and the higher resolution \htwo\
studies only appear amorphous in the outermost regions. 

Y1999 point out that dynamically there appear to be two types of CO emitters. The first type
is a group of small sources all found within 300\arcsec\ of the central star. The velocities
of these sources are distributed as if they all belong to an expanding torus region, which 
produces a nearly sinusoidal variation in the radial velocities with an amplitude of $\pm$17 
\kms , as found earlier by Healy \&\ Huggins (1990). They call these sources the inner ring.
The second type of emitters are found in the regions they call the outer arcs. 
The explanation for these two velocity systems was presented in OMM2004, who demonstrated that these 
are knots associated with the outer parts of the inner-disk and the outer-ring, which have different
expansion velocities and tilts with respect to the plane of the sky. 
The absence of CO emission from the PDR's surrounding the nebular ionization fronts is probably due 
to the fact that the optical depth in the photo-dissociating continuum doesn't become large
enough to allow formation of CO. Certainly, there is no evidence for a large optical depth in the
visual continuum on the outside of the nebula as there is no obvious depletion of stars.

\subsection{Conclusions}

Our observations and models of the knots in the Helix nebula have shown that these are objects strongly affected by the radiation field of the central star. The central densities of their cores are about
10$^{6}$ \cmq, with the side facing the star being photoionized. The peculiar surface brightness distribution of the [N~II] and \Ha\ cusps is explained by the process of photo-ablation of material from the core.
The knots are at the extreme of the regime of photoevaporation flows in terms of the importance of advection.
The \htwo\ emission arises from warm regions immediately behind the bright cusps, with a density of 
about 10$^{5}$ \cmq\ and with a temperature that cannot be explained by radiative heating and cooling.
The shadowed portions of the tails behind the cores are easily seen in CO because of the increased 
optical depth in the stellar continuum radiation. We point out that earlier calculations of conditions 
in the molecular zones around the PN have temperatures that are too high and that previous application of these models to observations of other PN were flawed.

\acknowledgments
We are grateful to Bruce Balick of the University of Washington and Arsen Hajian of the United States
Naval Observatory for their participation planning the observations reported on in this paper.
We also thank Robin J. R. Williams for many helpful conversations.
Our thanks also go to 
Angela K. Speck of the University of Missouri for clarification of the difference in meaning
of the surface brightness values for NGC 7293 included in S2002 and S2003 and for 
discussions on comparison of the observed average \htwo\ surface brightness with
the models of Vicini \etal\ (1999).
CRO's work on this project was partially supported by the STScI grant GO-9489.
WJH acknowledges financial support from DGAPA-UNAM, Mexico, through
project PAPIIT-IN115202 and through a sabbatical grant, and is also
grateful to the University of Leeds, UK, for hospitality during his
sabbatical visit.
GJF's work was supported in part by NASA NAG5-12020 and STScI AR 10316.

\clearpage

\end{document}